\documentclass[final,3p,times]{elsarticle}

\usepackage{amsmath,hyperref}
\usepackage{lineno}

\usepackage{epstopdf}
\journal{}

\usepackage{lineno,hyperref}
\usepackage{graphicx}
\usepackage{dcolumn}
\usepackage{bm}
\usepackage{epsfig}
\usepackage{booktabs}
\usepackage{subfigure}
\usepackage{graphics}
\usepackage{amssymb}
\usepackage{amsmath}
\usepackage{array}
\usepackage{color}
\usepackage{booktabs}
\usepackage{multirow}
\usepackage{caption}
\usepackage{chngpage}
\usepackage{subfigure}
\usepackage{mathrsfs,gensymb,float}

\usepackage[ruled,vlined,algo2e]{algorithm2e}

\biboptions{numbers,sort&compress}
\modulolinenumbers[1]

\begin{document}

\captionsetup[figure]{labelfont={bf},name={Fig.},labelsep=period}        

\begin{frontmatter}
	
	\title{Freezing dynamics of wetting droplet under a uniform electric field}


	\author[1,3,4]{Jiangxu Huang}
	\author[5]{Hanqing Li}
	\author[5]{Jiaqi Che}
	\author[1,3,4,6]{Zhenhua Chai\corref{mycorrespondingauthor}}	
	\cortext[mycorrespondingauthor]{Corresponding author}
	\ead{hustczh@hust.edu.cn}
	\author[2]{Lei Wang}
	\author[1,3,4]{Baochang Shi}

	\address[1]{ School of Mathematics and Statistics, Huazhong University of Science and Technology, Wuhan 430074, China}
	\address[3]{Institute of Interdisciplinary Research for Mathematics and Applied Science,Huazhong University of Science and Technology, Wuhan 430074, China}	
	\address[4]{Hubei Key Laboratory of Engineering Modeling and Scientific Computing, Huazhong University of Science and Technology, Wuhan 430074, China}	
	\address[5]{ Marine Design \& Research Institute of China}
	\address[6]{ The State Key Laboratory of Digital Manufacturing Equipment and Technology, Huazhong University of Science and Technology, Wuhan 430074, China}
	\address[2]{School of Mathematics and Physics, China University of Geosciences, Wuhan 430074, China}


\begin{abstract}

Electrofreezing is a powerful technique that employs the electric field to control and enhance the freezing process. In this work, a phase-field-based lattice Boltzmann (LB) method is developed to study the electrofreezing process of sessile droplet on a cooled substrate. The accuracy of the present LB method is first validated through performing some simulations of the three-phase Stefan problem, the droplet freezing on a cold wall, and the droplet deformation under a uniform electric field. Then it is used to investigate the effect of an electric field on the freezing of a wetting droplet on a cold substrate, and the numerical results show that the electric field has a significant influence on the freezing time of the droplet mainly through changing the morphology of the droplet. In particular, under the effect of the electric field, the freezing time is increased for the droplet with a prolate pattern, while the freezing time of the droplet with an oblate pattern is decreased. These numerical results bring some new insights on the electrofreezing and provide a valuable guidance for the precise regulation of droplet freezing.

\end{abstract}

\begin{keyword}
		Electrofreezing \sep freezing time \sep phase field \sep lattice Boltzmann method		
\end{keyword}
	
\end{frontmatter}

\section{Introduction}
The freezing of sessile droplets on the cold substrate is universal in both nature and industry. In the aerospace industry, the ice accumulation on aircraft wings can bring severe aerodynamic and flight mechanical effects, thus threatening aircraft flight safety \cite{PotapczukJAE2013}. In electric power systems, the ice accumulation in transmission lines would cause insulator flashovers \cite{LaforteAR1998}, and the icing on photovoltaic panels and wind turbines may reduce the production efficiency \cite{GaoIJHMT2019}. The processes mentioned above are closely related to the droplet freezing on a cold surface. Therefore, the suppression or elimination of droplet freezing on the cold surface is of great importance, which has also received increasing attention in recent years \cite{ZhaoIJHMT2020, TiwariATE2023}.

Over the past decades, some anti-icing strategies have been proposed to achieve freezing delay, which can be mainly divided into active and passive anti-icing technologies \cite{WangACSAMI2016}. The passive anti-icing technologies mainly rely on the use the superhydrophobic surface \cite{HeSM2010}, micro/nano-structured surface, and nanocoating \cite{HeSM2011, OberliACIS2014} to increase ice resistance. On the other hand, active anti-icing technologies usually require a continuous energy supply, including applying an external magnetic field \cite{GouETFS2024}, electric field \cite{DengIJTS2022, DengIECR2022}, and microwave radiation \cite{Dalvi-IsfahanJFE2017}. He et al. \cite{HeSM2010} experimentally investigated the freezing behavior of droplets on the hydrophobic and superhydrophobic thin films. The results show that compared to the hydrophobic surface, the frost formation on the superhydrophobic surface was delayed due to the freezing delay of liquid in the three-phase line region. Then, they \cite{HeSM2011} also fabricated the superhydrophobic surface with nanorod arrays and considered the freezing dynamics of microdroplets at both room and below-freezing temperatures. It is found that the superhydrophobic micro-droplets effectively retarding ice/frost formation at the below-freezing temperature. Oberli et al. \cite{OberliACIS2014} reviewed the condensation and freezing behaviors of droplets on the superhydrophobic surface, and demonstrated the superhydrophobic surface not only exhibits the excellent water repellency, but also reduces ice adhesion and delays water freezing. However, the surface may degrade during the thermal cycling or wetting. This degradation would further impair the superhydrophobicity and ice resistance of the superhydrophobic surface \cite{OberliACIS2014, VaranasiAPL2010}. In addition, the nanomaterials are expensive, and the thickness of the coating increases the thermal resistance of convection heat transfer of air, which in turn affects the efficiency of the device \cite{GouETFS2024}. Therefore, developing an effective anti-freezing method is an urgent issue that needs to be solved.

Since the first report of electrofreezing by Dufour \cite{DufourADP1862} in 1861, the electric field has been recognized as an effective technique for promoting nucleation, and has attracted extensive attention due to its advantages of the low power consumption, intelligent control, and no moving parts \cite{HeIJMF2023}. Actually, there are two main modes of applying electric fields: intrusive and non-intrusive. In the intrusive mode, the electrodes are in contact with the droplet and the substrate, generating an electric field inside the droplet \cite{CarpenterLangmuir2015}. In the non-intrusive mode, the electrode does not contact the droplet. Instead, the droplet is placed between electrode plates, creating an electric field around it. This non-intrusive mode seems to be more convenient and suitable for enhancing freezing \cite{AcharyaASIS2018}. Acharya and Bahadur \cite{AcharyaASIS2018} reviewed experimental studies and molecular dynamics simulations of ice nucleation under an applied electric field, and illustrated that the electric field can be used as a powerful tool to facilitate and control the freezing process. Recently, Deng et al. \cite{DengIJTS2022} experimentally investigated the effects of surface temperature and electric field strength on the freezing of sessile droplets on the superhydrophobic aluminum surface. They found that the electrostatic field promotes droplet nucleation, and the stronger electrostatic field enlarges the freezing time. Subsequently, Deng et al. \cite{DengIECR2022} also considered the effect of microstructure on droplet freezing under an electric field and reported that the applied electric field could be used to enhance the nucleation promotion and increase the freezing temperature, and the localized electric field enhancement occurred at the microstructure interface. From the above review, it is clear that experimental research is the main approach to understanding the effect of electric fields on droplet freezing. However, due to the challenges of experimental measurement, the existing studies primarily focus on the influence of electric field on the macroscopic dynamic behavior of droplet during the freezing process while overlooking many microscopic details, for instance, the interface dynamics. With the development of experimental techniques, some researchers have adopted molecular tagging thermometry (MTT) to achieve time- and space-resolved temperature distribution, revealing the time evolution of non-steady heat transfer and dynamic phase transition process inside microdroplets during freezing \cite{HuIJMF2010}. However, due to the highly non-uniform distribution of electric field during the freezing process control and the possibility of fluorescent molecules carrying a charge, there are some significant challenges for experimental works based on the MTT method.

With the development of computer technology and numerical methods, numerical simulation plays an essential role in the study of droplet freezing under an applied electric field. Molecular dynamics simulation has been widely used to study the freezing of liquid droplets under electric fields. However, the computational cost of this method is very high, and it is usually only used for nanoscale problems. In recent years, the lattice Boltzmann (LB) method based on the kinetic theory has great potential in the study of electrohydrodynamics \cite{LiuPOF2019, HeIJMF2023, HeATE2022, LuoPRE2019} and droplet freezing \cite{ZhangPRE2020, GuoIJTS2024, MohammadipourJFM2024, HuangFreezing2024}. Zhang et al. \cite{ZhangPRE2020} proposed an axisymmetric pseudo-potential LB model to simulate the freezing processes of sessile droplets, and considered the volume expansion of the droplets. The results show that the LB model can accurately describe the freezing processes of such sessile droplets. Based on this work \cite{ZhangPRE2020}, Guo et al. \cite{GuoIJTS2024} systematically studied the wetting droplet freezing on a cold wall, and mainly focused on the effect of the solid-liquid density ratio. Recently, Mohammadipour et al. \cite{MohammadipourJFM2024} developed a phase-field based LB model for the freezing of droplets, and the numerical results are in good agreement with the experimental data. Besides, they also investigated the effect of environmental conditions on the solidification of droplets. Through considering the density-induced volume change during droplet solidification, Huang et al. \cite{HuangFreezing2024} proposed an improved phase-field based LB model for simulating containerless freezing, and the numerical results agree well with both analytical and experimental data. Over the past two decades, numerical simulations of electrofreezing have primarily focused on the molecular dynamics simulations (MDS) at the nanoscale. Although these studies have provided significant insights into the electro-freezing process, it is usually difficult to validate the MDS-based results directly via the experiments due to the limitations of experimental conditions \cite{AcharyaACIS2018}. Thus it is desirable to conduct numerical simulations of electrofreezing at mesoscopic or macroscopic scales to overcome these limitations, providing results that are more experimentally verifiable and practically applicable. In this work, we will extend the previous work \cite{HuangFreezing2024} to develop a phase-field-based LB model for droplet freezing under an applied electric field.

The remainder of this paper is organized as follows. In section 2, we present the physical problem and governing equations for droplet freezing under an electric field, followed by the developed phase-field LB model. In Section \ref{sec4}, some numerical experiments are performed to test the present LB model, and the numerical results and discussion on the droplet freezing under an applied electric field are shown in Section \ref{sec5}. Finally, some conclusions are given in Section \ref{sec6}.

\section{Problem statement and governing equations}
\label{sec2}

\begin{figure}[H]
	\centering
	\includegraphics[width=0.45\textwidth]{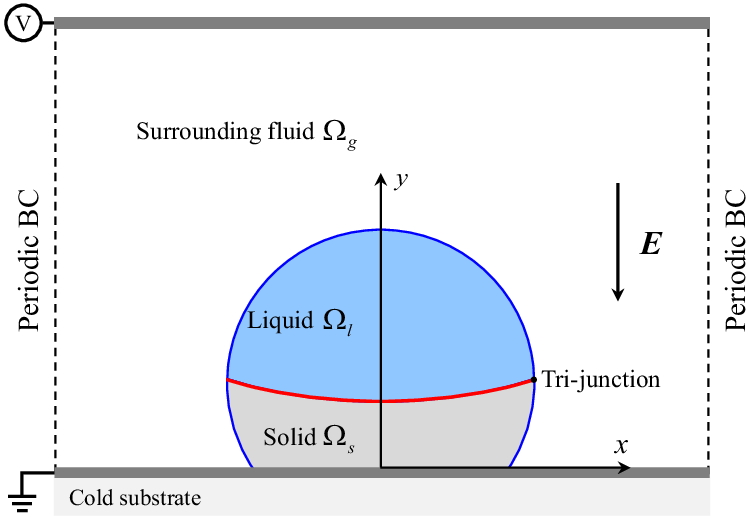}
	\caption{ Schematic of the droplet freezing on a cold substrate under an external electric field.}
	
	\label{fig1}
\end{figure}

Fig. \ref{fig1} illustrates a schematic diagram of the sessile droplet freezing in a uniform electric field. We assume that the solid phase only forms in the liquid phase, the liquid and gas phases are mutually immiscible, and the water vapor in the air is ignored. Initially, the whole region is kept at a constant temperature $T_g$, and a droplet with the radius of $R$ is placed on the bottom surface. Once the droplet reaches a specified contact angle $\theta$, a vertical electric field with the strength $\bm{E}$ is applied between the upper and lower electrodes, and a lower temperature $T_w$ is applied simultaneously to the bottom surface. The upper electrode plate with high potential will undergo complex electrochemical reactions, which will lead to the free charges into the system and further affect the freezing dynamics of the wetting droplet.

During the solidification process, the problem is a three-phase system involving complex changes of the phase interfaces. In this work, the droplet profile is captured mainly by the order parameter $\psi$, and the solid-liquid phase interface is tracked by the solid phase fraction $f_s$. The combination of the order parameter and solid fraction can be used to distinguish the three phases, i.e., $\psi=0$ and $f_s=0$ represent the gas phase, $\psi=1$ and $f_s=0$ denote the liquid phase, and $\psi=1$ and $f_s=1$ represent the solid phase. In this case, the following linear function of the order parameter and solid fraction can be used to characterize the material property of the system \cite{HuangFreezing2024},
\begin{equation}
	\chi=f_s \chi_s+\left(1-f_s\right) \psi \chi_l+\left(1-f_s\right)(1-\psi) \chi_g,
	\label{Computed}
\end{equation}
where placeholders $\chi$ characterize physical parameters, such as density, viscosity, thermal conductivity, dielectric constant, or conductivity, while subscripts $g$, $l$, and $s$ denote the gas, liquid, and solid phase region, respectively. In order to describe the interfacial evolution between the gas and liquid phases, the Allen-Cahn equation for the phase field and the incompressible Naiver-Stokes equations for the flow field are considered \cite{HuangFreezing2024},
\begin{subequations}
	\begin{equation}
		\frac{\partial \psi}{\partial t}+\nabla \cdot(\psi \mathbf{u})=\nabla \cdot[M(\nabla \psi-\lambda \mathbf{n})] + \psi \nabla \cdot\mathbf{u},
		\label{phasefield}
	\end{equation}

	\begin{equation}
		\nabla \cdot \mathbf{u}=(1-\frac{\rho_s}{\rho_l}) \frac{\partial f_s}{\partial t},
		\label{ns1}
	\end{equation}
	
	\begin{equation}
		\frac{\partial \rho \mathbf{u}}{\partial t}+\nabla \cdot(\rho \mathbf{u u})=-\nabla p+\nabla \cdot\left[\mu\left(\nabla \mathbf{u}+(\nabla \mathbf{u})^{\mathrm{T}}\right)\right]+\mathbf{F}_s+\mathbf{F}_b+\mathbf{F}_e +  \rho \mathbf{f},
		\label{ns2}
	\end{equation}
	\label{eq2}
\end{subequations}
where $\mathbf{u}$ is the velocity, $\rho$ is the density, $p$ is the pressure, $\mu$ is the dynamic viscosity, $M$ is a positive constant named mobility, $\lambda=4 \psi(1-\psi)/W$ with $W$ being the interface thickness. $\mathbf{F}_b$ is the body force,  $\mathbf{F}_e$ is the electrical force, $\mathbf{f}$ is the force for the fluid-solid interaction, $\mathbf{n}=\nabla \phi /|\nabla \phi|$ is the unit vector normal to the interface. The surface tension force $\mathbf{F}_s$ is define by \cite{LiuMMS2022}
\begin{equation}
	\mathbf{F}_s=\mu_\psi \nabla \psi,
	\label{Fs}
\end{equation}
where $\mu_\psi$ is the chemical potential defined by \cite{LiuMMS2022}
\begin{equation}
	\mu_\psi=4 \beta\left(\psi-\psi_l\right)\left(\psi-\psi_g\right)\left(\psi-\frac{\psi_l+\psi_g}{2}\right)-\kappa \nabla^2 \phi,
\end{equation}
where the parameters $\beta$  and $\kappa$  are related to the surface tension $\gamma$ and the interface thickness $W$ \cite{LiuMMS2022}, 
\begin{equation}
	k=\frac{3}{2} \gamma W, \quad \beta=\frac{12 \gamma}{W} .
\end{equation}
In addition, it should be noted that the source term $\dot{m}=\left(1-\frac{\rho_s}{\rho_l}\right) \frac{\partial f_s}{\partial t}$ on the right-hand side of the continuity equation (\ref{ns1}) arises from the volume change due to the density variation of the phase change material during the freezing process, and the details can be found in our previous work \cite{HuangFreezing2024}. 

The enthalpy-based energy equation is adopted to describe the freezing process, in which the enthalpy can be used to capture the evolution of the solid-liquid phase interface \cite{HuangFreezing2024}, 
\begin{equation}
	\frac{\partial \left(\rho H\right)}{\partial t}+\nabla \cdot\left(\rho C_p T \mathbf{u}\right)=\nabla \cdot(\lambda \nabla T) + \rho C_p T \nabla \cdot \mathbf{u},
	\label{eqTem}
\end{equation}
where $C_p$ is the specific heat capacity, $T$ is the temperature, $\lambda$ is the thermal conductivity, $H=C_p T+L f_l $ is the total enthalpy with $L$ being the latent heat. The temperature $T$ and the liquid phase rate $f_l$ can be uniquely determined by the total enthalpy $H$ \cite{LuoPRE2019, HuangIJHMT2013},  
\begin{equation}
	f_l =\left\{\begin{array}{ll}
		0 & H<H_s=C_p T \\
		\frac{H-H_s}{H_l-H_s} & H_s \leqslant H \leqslant H_l=H_s+L, \\
		1 & H>H_l
	\end{array} \quad T= \begin{cases}H / C_p & H<H_s \\
		T_s+\frac{H-H_s}{H_l-H_s}\left(T_l-T_s\right) & H_s \leqslant H \leqslant H_l, \\
		T_l+\left(H-H_l\right) / C_p & H>H_l\end{cases}\right.
	\label{Hfl}
\end{equation}
where $T_s$ and $T_l$ are the solidus and liquidus temperatures, respectively.

The electrohydrodynamic behavior of the droplet is described by the Taylor-Melcher leaky dielectric model \cite{TaylorMPS1966, MelcherARFM1969}. For the dielectric system considered in this work, the effect of magnetic induction is negligible due to the small dynamic current. Therefore, the electric field strength in the medium is spinless and can be expressed as
\begin{equation}
	\nabla \times \mathbf{E}=0.
\end{equation}
In addition, the Gauss' law for a dielectric fluid with dielectric constant $\varepsilon$ can be written as
\begin{equation}
	\nabla \cdot(\varepsilon \mathbf{E})=q_v,
\end{equation}
where $q_v$ is the volumetric free charge density. In addition, the electric field strength $\mathbf{E}$ can be expressed as a gradient of the electric potential, i.e.,  $\mathbf{E}=-\nabla \phi$. Then, the charge conservation equation is given by
\begin{equation}
	\frac{D q_v}{D t}+\nabla \cdot \mathbf{J}=\frac{\partial q_v}{\partial t}+\mathbf{u} \cdot \nabla q_v+\nabla \cdot(\sigma \mathbf{E})=0,
	\label{aa}
\end{equation}
where $\mathbf{J}=\sigma \mathbf{E}$ is the current density due to conductivity, $D / D t=\partial / \partial t+\mathbf{u} \cdot \nabla$ is the material derivative, and $\sigma$ is the electrical conductivity of the fluid. For the leaky dielectric model, the charge is confined to the surface, and the free charge in the bulk fluid is zero, i.e., $Dq_v/Dt=0$. Therefore, Eq. (\ref{aa}) can be rewritten as
\begin{equation}
	\boldsymbol{\nabla} \cdot(\sigma \mathbf{E})=0.
	\label{bb}
\end{equation}
Substituting $\mathbf{E}=-\nabla \phi$ into Eq. (\ref{bb}), one can obtain the governing equation for the electric field, 
\begin{equation}
	\boldsymbol{\nabla} \cdot(\sigma \nabla \phi)=0.
	\label{eqEle}
\end{equation}
Furthermore, the electrohydrodynamic phenomena are caused by electric stresses generated at interfaces, which can be represented by the divergence of the Maxwell stress tensor $\boldsymbol{\sigma}_M=\epsilon\left(\mathbf{E E}-\frac{1}{2}(\mathbf{E} \cdot \mathbf{E}) \mathbf{I}\right)$,
\begin{equation}
	\mathbf{F}_e=\nabla \cdot \boldsymbol{\sigma}_M=[\nabla \cdot(\epsilon \mathbf{E})] \mathbf{E}-\frac{1}{2}(\mathbf{E} \cdot \mathbf{E}) \nabla \epsilon+\frac{1}{2}(\mathbf{E} \cdot \mathbf{E}) \nabla\left(\rho \frac{\partial \epsilon}{\partial \rho}\right)_T
\end{equation}
where the three terms on the right-hand side of the above equation denote the Coulomb force acting along the electric field, dielectric force, and electrostrictive force. Following the previous works \cite{LiuPOF2019}, the electrostrictive force is usually neglected, and the electric force can be simplified by
\begin{equation}
	\mathbf{F}_e = q_v \mathbf{E}-\frac{1}{2} \mathbf{E} \cdot \mathbf{E} \nabla \mathcal{\varepsilon}.
	\label{Fe}
\end{equation}

In summary, the governing equations for the problem considered here are composed of Eqs (\ref{eq2}), (\ref{eqTem}), and (\ref{eqEle}), which will be solved by different LB schemes shown in the next section.

\section{Numerical methods}
\label{sec3}

\subsection{LB model for the phase field}
Similar to the previous LB models for the Allen-Cahn equations, the LB evolution equation with the BGK collision operator for the phase field can be expressed as \cite{ChaiJSC2016, LiangPRE2018}
\begin{equation}
	f_i\left(\mathbf{x}+\mathbf{c}_i \Delta t, t+\Delta t\right)-f_i(\mathbf{x}, t)=-\frac{1}{\tau_f}\left[f_i(\mathbf{x}, t)-f_i^{\mathrm{eq}}(\mathbf{x}, t)\right]+ \left(1-\frac{1}{2 \tau_f}\right) \Delta t F_i(\mathbf{x}, t),
	\label{LB-PF}
\end{equation}
where $f_i(\mathbf{x}, t)$ is the distribution function at position $\mathbf{x}$ and time $t$, $\mathbf{c}_i$ is the discrete velocity, , $\Delta t$ is the time step. $\tau_f=M/c_s^2 \Delta t + 0.5$ is the relaxation time for phase field with $c_s = c /\sqrt 3$ being the sound speed. For the two-dimensional LB method considered here, the nine-velocity (D2Q9) lattice model is adopted, the weight coefficients $\omega_i$ and discrete velocities $\mathbf{c}_i$ are defined as
\begin{equation}
	\omega_i= \begin{cases}4 / 9 & i=0, \\ 1 / 9 & i=1-4, \\ 1 / 36 & i=5-8,\end{cases}
\end{equation}
\begin{equation}
	\mathbf{c}_i= \begin{cases}(0,0), & i=0, \\ (\cos [(i-1) \pi / 2], \sin [(i-1) \pi / 2]) c, & i=1-4, \\ (\cos [(2 i-9) \pi / 4], \sin [(2 i-9) \pi / 4]) \sqrt{2} c, & i=5-8,\end{cases}
\end{equation}
where $c = \Delta x/\Delta t$ is the lattice speed with $\Delta x$ denoting the lattice spacing (for simplicity, $\Delta x=\Delta t=1$ is considered). The local equilibrium distribution function $f_i^{\mathrm{eq}}$ and the source term $F_i$ are defined by \cite{ChaiJSC2016}
\begin{equation}
	f_i^{\mathrm{eq}}=\omega_i \psi\left(1+\frac{\mathbf{c}_i \cdot \mathbf{u}}{c_s^2}\right),
\end{equation}
\begin{equation}
	F_i= \frac{\omega_i \mathbf{c}_i \cdot\left[\partial_t(\psi \mathbf{u})+c_s^2 \lambda \mathbf{n}\right]}{c_s^2} + \omega_i \psi \nabla \cdot\mathbf{u}.
	\label{eq23}
\end{equation}
In addition, the order parameter is calculated by
\begin{equation}
	\psi=\sum_i f_i + \frac{\Delta t}{2} \psi \nabla \cdot\mathbf{u}.
	\label{Computed_psi}
\end{equation}

\subsection{LB model for the flow field}

Compared to the previous works for incompressible fluid flows, there is a source term in the continuity equation, which is used to describe the freezing process with the volume change. Here, the generalized LB model proposed by Yuan et al. \cite{YuanCMA2020} is considered, and the evolution can be written as
\begin{equation}
	\begin{aligned}
		h_i\left(\mathbf{x}+\mathbf{c}_i \Delta t, t+\Delta t\right)-h_i(\mathbf{x}, t)=  -\frac{1}{\tau_h}\left[h_i(\mathbf{x}, t)-h_i^{\mathrm{eq}}(\mathbf{x}, t)\right] +\Delta t \left(1-\frac{1}{2 \tau_h}\right)  R_i(\mathbf{x}, t),
	\end{aligned}
	\label{LB_flow}
\end{equation}
where $\tau_h = \nu / {c_s}^2 dt +0.5$ is the corresponding relaxation time for flow field , $h_i^{\mathrm{eq}}$ is the local equilibrium distribution function for flow field \cite{YuanCMA2020},
\begin{equation}
	\begin{aligned}
		&h_i^{\mathrm{eq}}= \begin{cases}\frac{p}{c_s^2}\left(\omega_i-1\right)+\rho s_i(\mathbf{u}), & \mathrm{i}=0 \\ \frac{p}{c_s^2} \omega_i+\rho s_i(\mathbf{u}), & \mathrm{i} \neq 0\end{cases}
	\end{aligned}
\end{equation}
with
\begin{equation}
	\begin{aligned}
		s_i(\mathbf{u})=\omega_i\left[\frac{\mathbf{c}_i \cdot \mathbf{u}}{c_s^2}+\frac{\left(\mathbf{c}_i \cdot \mathbf{u}\right)^2}{2 c_s^4}-\frac{\mathbf{u} \cdot \mathbf{u}}{2 c_s^2}\right] .
	\end{aligned}
\end{equation}
The forcing term $R_i$ can be expressed as \cite{YuanCMA2020, HuangFreezing2024}
\begin{equation}
	R_i=\omega_i  \left[S +\frac{\mathbf{c}_i \cdot (\mathbf{F}+\rho \mathbf{f})}{c_s^2} + \frac{(\mathbf{u}\tilde{\mathbf{F}} +\tilde{\mathbf{F}} \mathbf{u}):\left(\mathbf{c}_i \mathbf{c}_i-c_s^2 \mathbf{I}\right)}{2c_s^4}\right],
\end{equation}
where $ S=\rho \dot{m} + \mathbf{u} \cdot \nabla \rho $, $ \tilde{\mathbf{F}} = \mathbf{F} - \nabla p + c_s^2 \nabla \rho +  c_s^2 \nabla \cdot S $, $\mathbf{F}=\mathbf{F}_s+\mathbf{F}_b+\mathbf{F}_e$ is the total force. The macroscopic quantities of flow field, i.e.,  $\mathbf{u}$ and $p$, can be evaluated by \cite{YuanCMA2020, HuangFreezing2024}
\begin{equation}
	\rho \mathbf{u}^*=\sum \mathbf{c}_i h_i+\frac{\Delta t}{2}  \mathbf{F},
\end{equation}
\begin{equation}
	\mathbf{u}=\mathbf{u}^*+\frac{\Delta t}{2} \mathbf{f},
	\label{Computed_u}
\end{equation}
\begin{equation}
	p=\frac{c_s^2}{\left(1-\omega_0\right)}\left[\sum_{i \neq 0} h_i+\frac{\Delta t}{2}S + \tau \Delta t R_0 + \rho s_0(\mathbf{u})\right],
	\label{Computed_p}
\end{equation}
where $\mathbf{u}^*$ is the velocity without considering the solid-liquid interaction, the force $\mathbf{f}$ used to reflect fluid-solid interaction $f_s \left(\mathbf{u}_s-\mathbf{u}^*\right) / \delta t$. We note that this approach for solid-liquid interaction has been widely applied to deal with some problems involving complex fluid-solid interfaces, such as particulate flows \cite{LiuCF2022}, dendrite growth \cite{ZhanAR2022} and flow in complex geometries \cite{ZhanPD2024}.

\subsection{LB model for the temperature field}

For the temperature field, the enthalpy-based thermal lattice Boltzmann model proposed by Huang et al. \cite{HuangIJHMT2013} is adopted to determine the total enthalpy distribution function $g_i(\mathbf{x}, t)$,
\begin{equation}
	g_i\left(\mathbf{x}+\mathbf{e}_i \Delta t, t+\Delta t\right)=g_i(\mathbf{x}, t)-\frac{1}{\tau_g}\left[g_i(\mathbf{x}, t)-g_i^{e q}(\mathbf{x}, t)\right]+ \left(1-\frac{1}{2 \tau_h}\right) \Delta t \rho C_p T \dot{m},
	\label{LB_H}
\end{equation}
where $\tau_h = \lambda / \rho C_{p, \mathrm{ref}} c_s^2 \Delta t$ is the relaxation time related to the thermal conductivity, $g_i^{e q}$ is the local equilibrium distribution function defined as \cite{HuangIJHMT2013}
\begin{equation}
	g_i^{e q} = \begin{cases}H-C_{p, \text { ref }} T+\omega_i C_p T\left(\frac{C_{p, \mathrm{ref}}}{C_p}-\frac{\mathbf{I}: \mathbf{u} \mathbf{u}}{2 c_s^2}\right), & i=0, \\ \omega_i C_p T\left[\frac{C_{p, \text { ref }}}{C_p}+\frac{\mathbf{e}_i \cdot \mathbf{u}}{c_s^2}+\frac{\left(\mathbf{e}_i \mathbf{e}_i-c_s^2 \mathbf{I}\right): \mathbf{u} \mathbf{u}}{2 c_s^4}\right], & i \neq 0,\end{cases}
\end{equation}
where $C_{p, \mathrm{ref}}=2 C_{p, s} C_{p, l} /\left(C_{p, s}+C_{p, l}\right)$ is the reference specific heat capacity \cite{HuangIJHMT2013}. The total enthalpy is calculated as follows,
\begin{equation}
	H=\sum_{i=0} g_i + \frac{1}{2} \Delta t \rho C_p T \dot{m}.
\end{equation}
Once the enthalpy is updated by the above equation, the liquid fraction and temperature can be determined by Eq. (\ref{Hfl}).

\subsection{LB model for the electric field}
As a special form of the Poisson equation, the Laplace equation (\ref{eqEle}) for the electric field can also be solved in the framework of LB method \cite{HirabayashiJSME2001, HeCPC2000, GuoTJCP2005, WangJCIS2006}. Here, we adopt the LB model proposed by Chai et al. \cite{ChaiPLA2007, ChaiAMM2008, ChaiSIAMJSC2019} to solve the electric field because it can correctly recover to the Laplace equation. The LB evolution equation for the electric field can be expressed as \cite{LiuPOF2019, ChaiAMM2008}
\begin{equation}
	l_i\left(\boldsymbol{x}+c_i \Delta t, t+\Delta t\right)-l_i(\boldsymbol{x}, t)=-\frac{1}{\tau_l}\left(l_i(\boldsymbol{x}, t)-l_i^{e q}(\boldsymbol{x}, t)\right),
	\label{LB_phi}
\end{equation}
where $\tau_l = \sigma/c_s^2 \Delta t + 0.5$ is the relaxation time for the electric field, the local distribution function $l_i^{e q}$ is defined by \cite{LiuPOF2019, ChaiAMM2008}
\begin{equation}
	l_i^{e q}= \begin{cases}\left(\omega_0-1\right) \phi, & i=0, \\ \omega_i \phi, & i=1-4 .\end{cases}
\end{equation}
Considering that the linear equilibrium distribution function is used here, the simple two-dimensional five-velocity (D2Q5) lattice model is adopted, in which the corresponding discrete velocities $\mathbf{c}_i$ and weight coefficients $\omega_i$ are expressed as
\begin{equation}
	\mathbf{c}_i=\left[\begin{array}{rrccr}
		0 & 1 & 0 & -1 & 0 \\
		0 & 0 & 1 & 0 & -1
	\end{array}\right] c, \quad \omega_i= \begin{cases}\frac{1}{3}, & i=0, \\
		\frac{1}{6}, & i=1-4.\end{cases}
\end{equation}
The macroscopic electric potential $ \phi $ can be obtained by \cite{LiuPOF2019,ChaiAMM2008} 
\begin{equation}
	\phi=\sum_{i=1}^4 \frac{1}{1-\omega_0} l_i.
	\label{Computed_phi}
\end{equation}

In addition, the derivative terms in the LB model should be discretized with some suitable difference schemes. For simplicity, the temporal derivative in Eq. (\ref{eq23}) is calculated by the first-order explicit Euler scheme \cite{LiangPRE2018},
\begin{equation}
	\partial_t(\phi \mathbf{u})(\mathbf{x}, t)=[(\phi \mathbf{u})(\mathbf{x}, t)-(\phi \mathbf{u})(\mathbf{x}, t-\Delta t)] / \Delta t .
\end{equation}
The gradient and Laplace operators are approximated by the second-order isotropic central schemes \cite{LiuMMS2022, LiangPRE2018},
\begin{subequations}
	\begin{equation}
		\nabla \zeta(\mathbf{x}, t)  =\sum_{i \neq 0} \frac{\omega_i \mathbf{c}_i \zeta\left(\mathbf{x}+\mathbf{c}_i \Delta t, t\right)}{c_s^2 \Delta t},
		\label{eqtidi1}
	\end{equation}	
	\begin{equation}
		\nabla^2 \zeta(\mathbf{x}, t)  =\sum_{i \neq 0} \frac{2 \omega_i \left[\zeta\left(\mathbf{x}+\mathbf{c}_i \Delta t, t\right)-\zeta(\mathbf{x}, t)\right]}{c_s^2 \Delta t^2}.
		\label{eqtidi2}
	\end{equation}
\end{subequations}

Electrofreezing of a droplet is a multiple field-coupling problem involving phase, electric, flow, and temperature fields. To facilitate the implementation of the procedure, the details are shown in the following Algorithm \ref{alg1}, where the specific treatments on the boundary conditions are also specified.

\begin{algorithm2e}[H]
	\SetAlgoLined
	\DontPrintSemicolon
	\SetKwInOut{Input}{\textbf{Input}}
	\SetKwInOut{Output}{\textbf{Output}}	
	
	\SetKwProg{In}{Initialization}{}{}
	\SetKwProg{Main}{Main Iteration Loop}{}{}
	
	\In{}
	{
		Initialize the distribution functions $(f_i, g_i, h_i, l_i)$ for different physical fields;
	}
	\textbf{end}
	
	\Main{}
	{
		Implement phase-field LB equation via Eq. (\ref{LB-PF});
		
		Determine the unknown distribution functions of order parameters at boundary nodes via a half-way bounce-back scheme;
		
		Implement wetting boundary conditions \cite{LiangPRE2018-CA};
		
		Calculate the order parameters $\psi$, density $\rho$, and surface tension force $\mathbf{F}_s$ via Eqs. (\ref{Computed_psi}), (\ref{Computed}) and (\ref{Fs});
		
		Implement LB equation for the electric potential via Eq. (\ref{LB_phi});
		
		Determine the unknown distribution functions of electric potential at boundary nodes via the general bounce-back scheme \cite{ZhangPRE2012};
		
		Calculate the electric potential $\phi$ and electrical forces $\mathbf{F}_e$ via Eqs. (\ref{Computed_phi}) and (\ref{Fe});
		
		\While{Electric field is convergent}
		{

			Implement LB equation for the flow field via Eq. (\ref{LB_flow});
			
			Boundary conditions are treated by the half-way bounce-back scheme;
			
			Calculate the velocity $\mathbf{u}$ and pressure $p$ via Eqs. (\ref{Computed_u}) and (\ref{Computed_p});    
			
			Implement LB equation for total enthalpy
			via Eq. (\ref{LB_H};
			
			Determine the unknown distribution functions of total enthalpy at boundary nodes via the general bounce-back scheme \cite{ZhangPRE2012};
			
			Calculate the temperature $T$ and the liquid phase fraction $f_s$ via Eq. (\ref{Hfl});    	
			
		}
		
		\If{Output required}
		{
			Output global parameters file;
		}
	}
	\textbf{end}	
	
	\caption{Overview of the LB algorithm structure for the sessile droplet freezing in an applied electric field.}
	\label{alg1}
	
\end{algorithm2e}

\section{Model validation}
\label{sec4}

In this section, several numerical experiments are conducted to demonstrate the capability and reliability of the present LB method. The problems considered here include the three-phase Stefan problem, sessile droplet freezing, and deformation of a droplet in the electric field. The numerical results are also compared with the corresponding numerical, analytical, and experimental data reported in the available works.

\begin{figure}[H]
	\centering
	\includegraphics[width=0.3\textwidth]{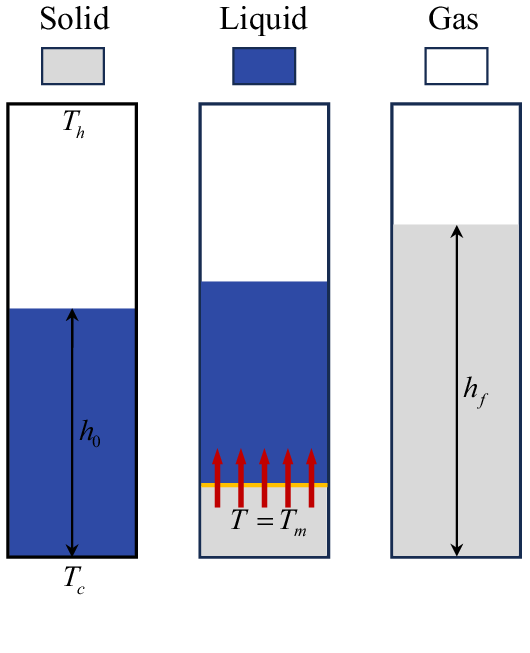} 
	\label{fig2a}
	\put(-160,165){\small (\textit{a})}
	\quad
	\quad
	\quad
	\includegraphics[width=0.45\textwidth]{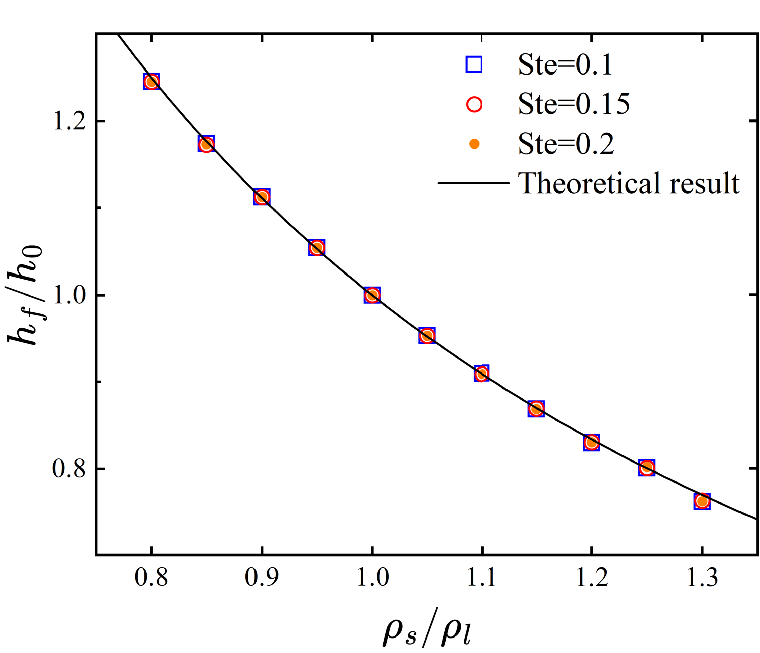} 
	\label{fig2b}
	\put(-210,165){\small (\textit{b})}
	
	\caption{Schematic diagram of the three-phase Stefan problem (a) and a comparison of volume change rates between the numerical and theoretical solutions under different Stefan numbers and solid-liquid density ratios (b).}
	\label{fig2}
\end{figure}

Firstly, to test the accuracy of the present LB method in the study of the volume expansion and contraction caused by density change during the freezing process, we perform some simulations of the three-phase Stefan problem. Initially, the cavity is filled with a liquid phase with the height $h_0$, and the rest of the region is filled with the gas phase and the temperature $T_h$ [Fig. \ref{fig2}(a)]. At $t>0$, a low temperature $T_w$ is applied to the bottom of the cavity, and the liquid phase begins to freeze into a solid phase. After the complete freezing, the height of the solidified liquid is equal to $h_f$. According to mass conservation, the final height of the solid phase can be obtained from the relation $h_f=\rho_l h_0 / \rho_s $ \cite{ShetabivashJCP2020}. In our simulations, the periodic boundary conditions are applied to the left and right boundaries, and the bounce-back scheme is used to treat the boundary conditions imposed on the top and bottom boundaries.

Fig. \ref{fig2}(b) presents a comparison between the numerical results and the theoretical solution of the final solid phase height $h_f/h_0$ under different values of the solid-liquid density ratio $\rho_s/ \rho_l$ and  Stefan number $\mathit{Ste}$ ($\mathit{Ste}={C_p\left(T_m-T_w\right)}/{L}$). The results show that the present LB method can accurately capture the volume change during the freezing process.

\begin{figure}[H]
	\centering
	\includegraphics[width=0.9\textwidth]{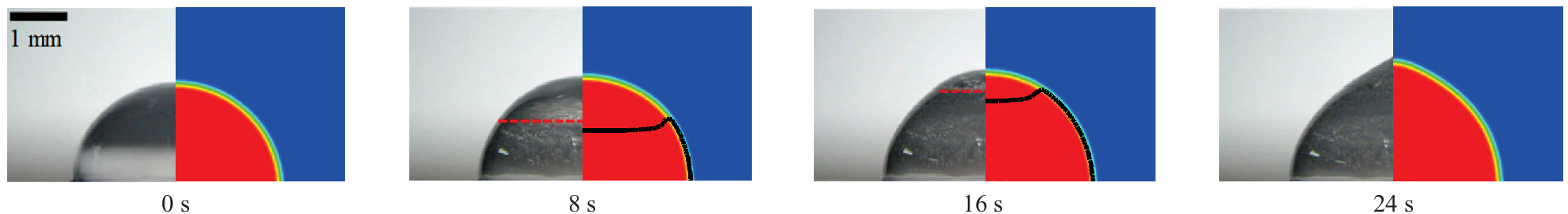} 
	\label{fig3a}
	\put(-440,50){\small (\textit{a})}
	\vspace{0.15cm}
	\includegraphics[width=0.9\textwidth]{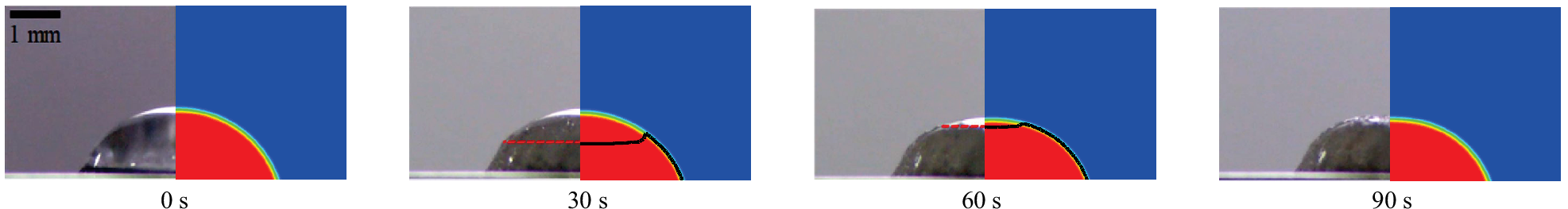} 
	\label{fig3b}
	\put(-440,50){\small (\textit{b})}
	
	\caption{The comparison of experimental \cite{GuoIJTS2024} and numerical results on the droplet profiles and phase interfaces during the freezing process of water (a) and hexadecane (b) droplets on cold wall surfaces.}
	\label{fig3}
\end{figure}

\begin{figure}[H]
	\centering
	\label{key}	\includegraphics[width=0.45\textwidth]{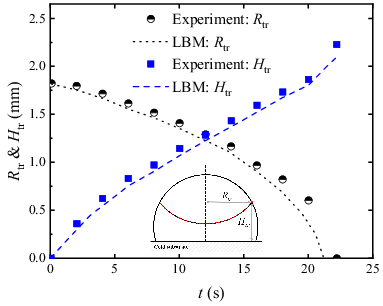} 
	\label{fig4a}
	\put(-210 ,170){\small (\textit{a})}
	\quad
	\includegraphics[width=0.45\textwidth]{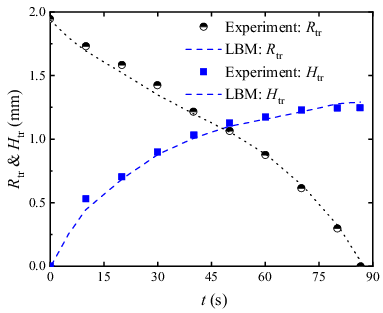} 
	\label{fig4b}
	\put(-210,170){\small (\textit{b})}
	
	\caption{The comparisons of the radius and height of the triple contact line between the present numerical and experimental results \cite{GuoIJTS2024}: water (a) and hexadecane (b).}
	\label{fig4}
\end{figure}

In order to show the capacity of the present LB method for simulating the freezing of wetting droplets, here we consider the problems in the previous work \cite{GuoIJTS2024}. In the experimental study \cite{GuoIJTS2024}, the $12 \ \mu \mathrm{L}$ water droplet and $10 \ \mu \mathrm{L}$ hexadecane droplet are placed on the solid wall with the temperatures of $-10 \ { }^{\circ} \mathrm{C}$ and $15 \ { }^{\circ} \mathrm{C}$, respectively. The droplet quickly begins to freeze once it touches the wall, and the supercooling stage is ignored. According to the physical parameters of water and hexadecane provided by Guo et al. \cite{GuoIJTS2024} (see Table \ref{tab1}), one can obtain the values of some dimensionless parameters, i.e., $\mathit{Ste}=0.13, \rho_s/ \rho_l=0.92, \theta=87^{\circ} $. We conduct some simulations and plot the results in Fig \ref{fig3}, where a comparison between the present numerical and the experimental results of the droplet profiles and the solid-liquid phase interfaces at different times is shown. From this figure, one can see that the freezing front gradually moves from the bottom to the top during the freezing process and forms a concave shape, which is consistent with that in Ref. \cite{MarinPRL2014}. The water droplet expands in volume after freezing and finally forms a conical tip at the top of the droplet. However, for the hexadecane droplet, it shrinks inwards during freezing, and eventually forms a platform at the top of the droplet. It is clear that the present LB method is able to capture the freezing characteristics of the droplet with different density ratios. In addition, Fig. \ref{fig4} presents a quantitative comparison between the numerical results and experimental data of the triple contact line radius $R_{tr}$ and height $H_{tr}$ of water and hexadecane droplets during the freezing processes. The results show that the present LB method is accurate in the study of the freezing dynamics of wetting droplets.

\begin{table}[H]
	\centering
	\caption{ Physical properties of water and hexadecane \cite{GuoIJTS2024}.}	
	\begin{tabular}{cccccc}
		\hline \hline
		Parames                          & Units                                           & \multicolumn{2}{c}{Water} & \multicolumn{2}{c}{Hexadecane} \\ \cline{3-6} 
		&                                                 & Liquid       & Solid      & Liquid         & Solid         \\ \hline
		Density $\rho$                   & $\mathrm{kg}\ \mathrm{m}^{-3}$                  & 999          & 917        & 774            & 833           \\
		Thermal conductivity $\lambda$   & $\mathrm{W} \mathrm{m}^{-1} \mathrm{~K}^{-1}$   & 0.581        & 2.16       & 0.15           & 0.15          \\
		Specific heat capacity $C_p$     & $\mathrm{J}\ \mathrm{kg}^{-1} \mathrm{~K}^{-1}$ & 4220         & 2100       & 2310           & 1800          \\
		Viscosity $\mu$                  & $\mathrm{Pa}\ \mathrm{s}$                       & 0.003        & -          & 0.003          & -             \\
		Surface tension $\sigma$         & $\mathrm{N}\ \mathrm{m}^{-1}$                   & 0.076        & -          & 0.028          & -             \\
		Latent heat $L$                  & $\mathrm{kJ}\ \mathrm{kg}^{-1}$                 & 333.4        & -          & 230            & -             \\
		Solidification temperature $T_m$ & ${ }^{\circ} \mathrm{C}$                        & 0            & -          & 18             & -             \\ \hline \hline
	\end{tabular}
	\label{tab1}
\end{table}

To further test the efficiency of the present LB method for simulating electrohydrodynamics (EHD) problems, a numerical study was conducted on a single dielectric droplet suspended in another leaky dielectric fluid under an applied electric field. Under the action of an external electric field, the interaction of electric stress and surface tension  force causes the droplet to form different deformation patterns and flow states, which are controlled by the following two dimensionless  parameters,
\begin{equation}
	D=\frac{L-H}{L+H}, \quad \zeta=\frac{R}{S},
	\label{D}
\end{equation} 
where $D$ is the deformation factor, $\zeta$ is the ratio of charge relaxation times between the internal and external phases, $L$ and $H$ are the end-to-end length of the stable droplet measured along and perpendicularly to the direction of the electric field, respectively. $R=\sigma_i / \sigma_e$ and $S=\varepsilon_i / \varepsilon_e$ are the 
conductivity and permittivity ratios of the internal and external phases of the droplet. Based on the Taylor's small deformation theory \cite{TaylorMPS1966}, the steady state deformation factor of a droplet in an electric field can be expressed as
\begin{equation}
	D=\frac{9 C a_E}{16} \frac{f_T(R, S, B)}{(2+R)^2},
	\label{taylor}
\end{equation}
where $f_T(R, S, B)=R^2+1-2 S+\frac{3}{5}(R-S) \frac{2+3 B}{1+B}$ is the function related to the conductivity ratio $R$, permittivity ratio $S$ and the viscosity ratio $B=\mu_i / \mu_e$. $C a_E=\varepsilon_e E_0^2 R / \gamma$ is the electric capillary number that characterizes the strength of the electric field. Based on Taylor’s small deformation theory, Feng et al.  \cite{FengJCIS2002} derived a first-order approximate expression for the deformation of a two-dimensional droplet in an electric field,
\begin{equation}
	D=\frac{C a_E f_F(R, S)}{3(1+R)^2},
\end{equation}
where $f_F(R, S)=R^2+R+1-3 S$ is a discriminant function that determines whether the droplet is deformed into a prolate or oblate shape.

\begin{figure}[H]
	\centering
	\includegraphics[width=0.45\textwidth]{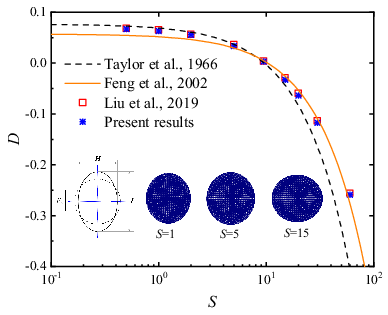} 
	\label{fig5a}
	\put(-210 ,170){\small (\textit{a})}
	\quad
	\includegraphics[width=0.45\textwidth]{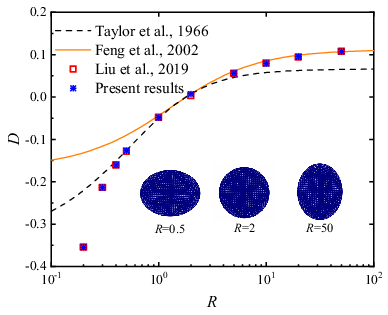} 
	\label{fig5b}
	\put(-210,170){\small (\textit{b})}
	
	\caption{A comparison of the deformation factors among the present results, the referenced numerical results \cite{LiuPOF2019} and the theoretical solutions \cite{TaylorMPS1966, FengJCIS2002}. $R=5.0$ and $Ca_E=0.2$ (a); $S=2.0 $ and $Ca_E=0.2$ (b).}
	\label{fig5}
\end{figure}

We carry out some simulations and plot the results in Fig. \ref{fig5}. From this figure, one can observe that the present numerical results are close to the theoretical solutions of Taylor \cite{TaylorMPS1966} and Feng \cite{FengJCIS2002}. In addition, as shown in Fig. \ref{fig5}, the droplet gradually changes from a prolate shape to an oblate shape as $S$ increase and $R$ decrease, and the present results agree well with Taylor's results when $R$ or $S$ is relatively small, while they agree well with Feng's results when $S$ and $R$ are large. Furthermore, we also note that our results are in good agreement with the numerical results reported in Ref. \cite{LiuPOF2019}.

\section{Numerical results and discussion}
\label{sec5}

\begin{figure}[H]
	\centering
	\includegraphics[width=0.5\textwidth]{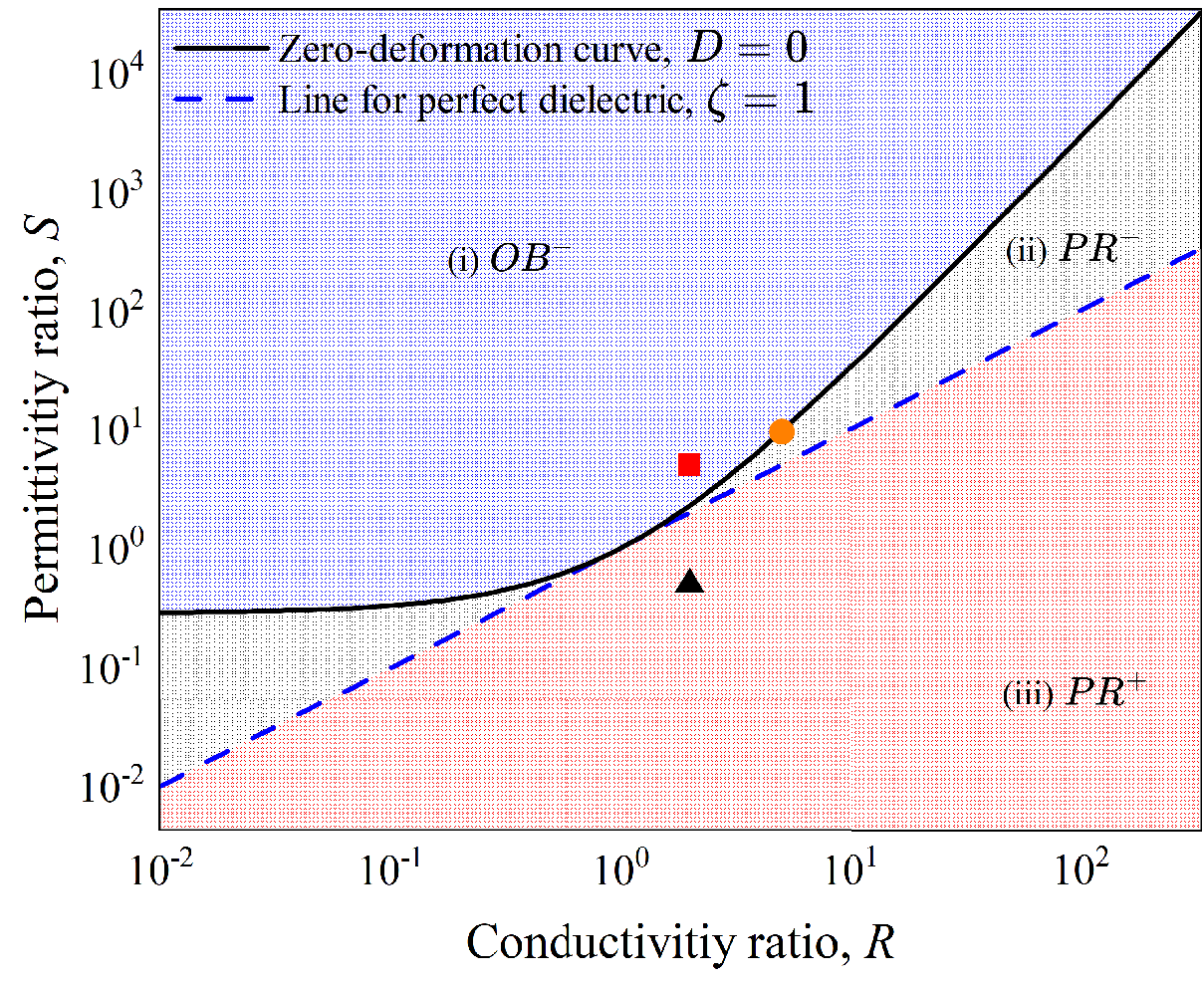} 
	\label{fig6a}
	\put(-234 ,187){\small (\textit{a})}
	\quad
	\quad
	\includegraphics[width=0.194\textwidth]{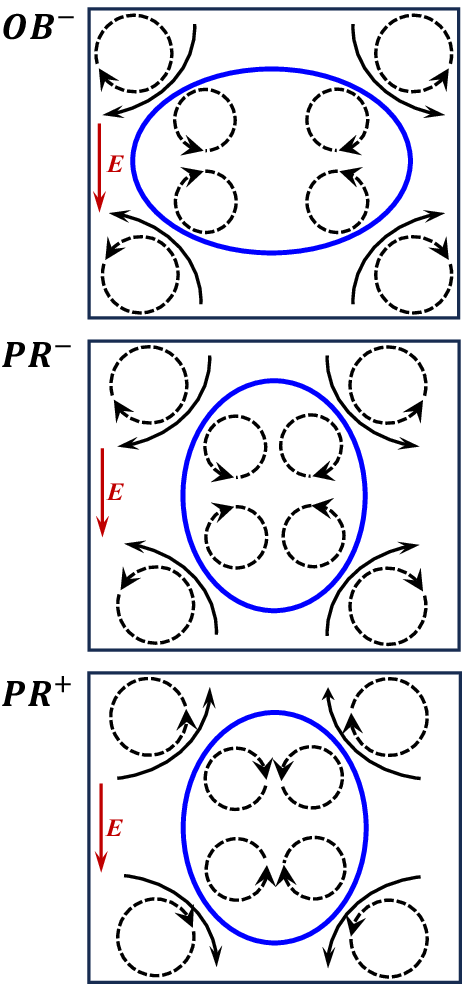} 
	\label{fig6b}
	\put(-108,187){\small (\textit{b})}
	
	\caption{$R-S$ space diagram of the equilibrium droplet shape and circulating flow pattern at the viscosity ratio of unity, the three symbols represent the three cases considered in this work (a); the droplet shapes and flow patterns of three different cases, the solid blue line shows the droplet profile (b).}
	\label{fig6}
\end{figure}

In this section, the effect of the electric field on the freezing dynamic behavior of a wetting droplet is investigated. According to Eqs. (\ref{D}) and (\ref{taylor}), we can plot the $R-S$ phase diagram of the droplet shape as well as the induced flow direction in Fig. \ref{fig6}(a). It can be seen from this figure that the droplet under the action of the electric field will show three typical states: (i) $OB^-$ is oblate droplet with the pole-to-equator induced flow, (ii) $PR^+$ is prolate droplet with the pole-to-equator induced flow, and (iii) $PR^-$ is prolate droplet with the equator-to-pole induced flow. Fig. \ref{fig6}(b) shows the droplet shapes and flow states of three typical cases corresponding to $OB^-: (R, S)=(2.0,5.0)$, $PR^-: (R, S)=(5.0,9.5)$ and $PR^+: (R, S)=(2.0,0.5) $. However, the freezing process of a droplet on the cooled substrate can be divided into four different cases based on thermomechanical characteristics: supercooling, crystallization nucleation, equilibrium freezing, and solid subcooling \cite{AkhtarRSER2023}. This is a multi-scale process with the difference in the time scales between the crystallization nucleation phase and the solidification phase being over three orders of magnitude \cite{AkhtarRSER2023}. Thus, the simulation of ice crystal growth at full scale is very difficult, as the computational cost is very expensive, even with a high-performance computer. For this reason, the crystallization nucleation stage has often been neglected in the previous studies \cite{ZhangPRE2020, GuoIJTS2024, MohammadipourJFM2024}, and the period before the nucleation phase is always considered as the initial state in numerical simulations.      

\subsection{Effect of electric field on the evolution of freezing droplet morphology }
\begin{figure}[H]
	\centering
	\includegraphics[width=0.2\textwidth]{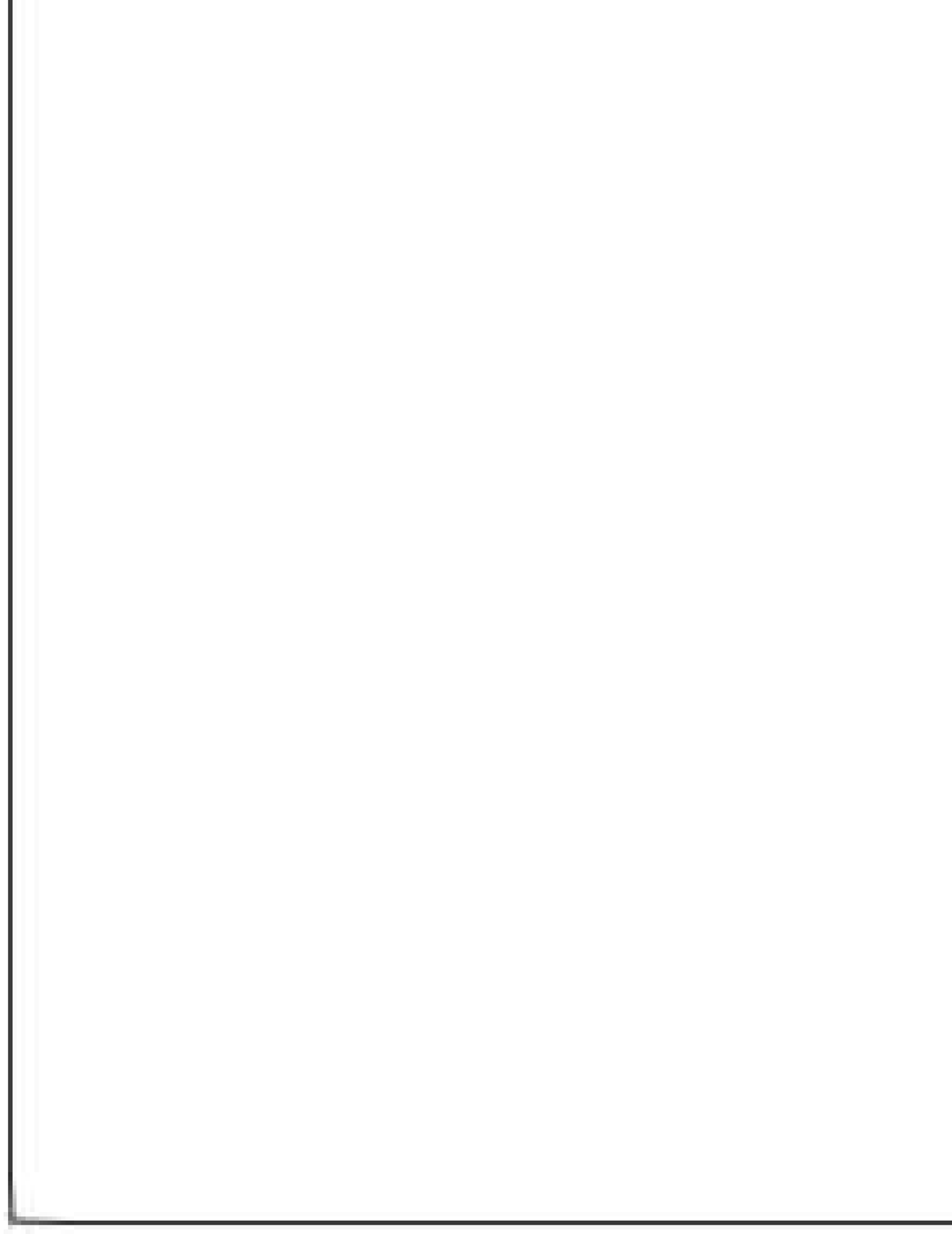} 
	\includegraphics[width=0.2\textwidth]{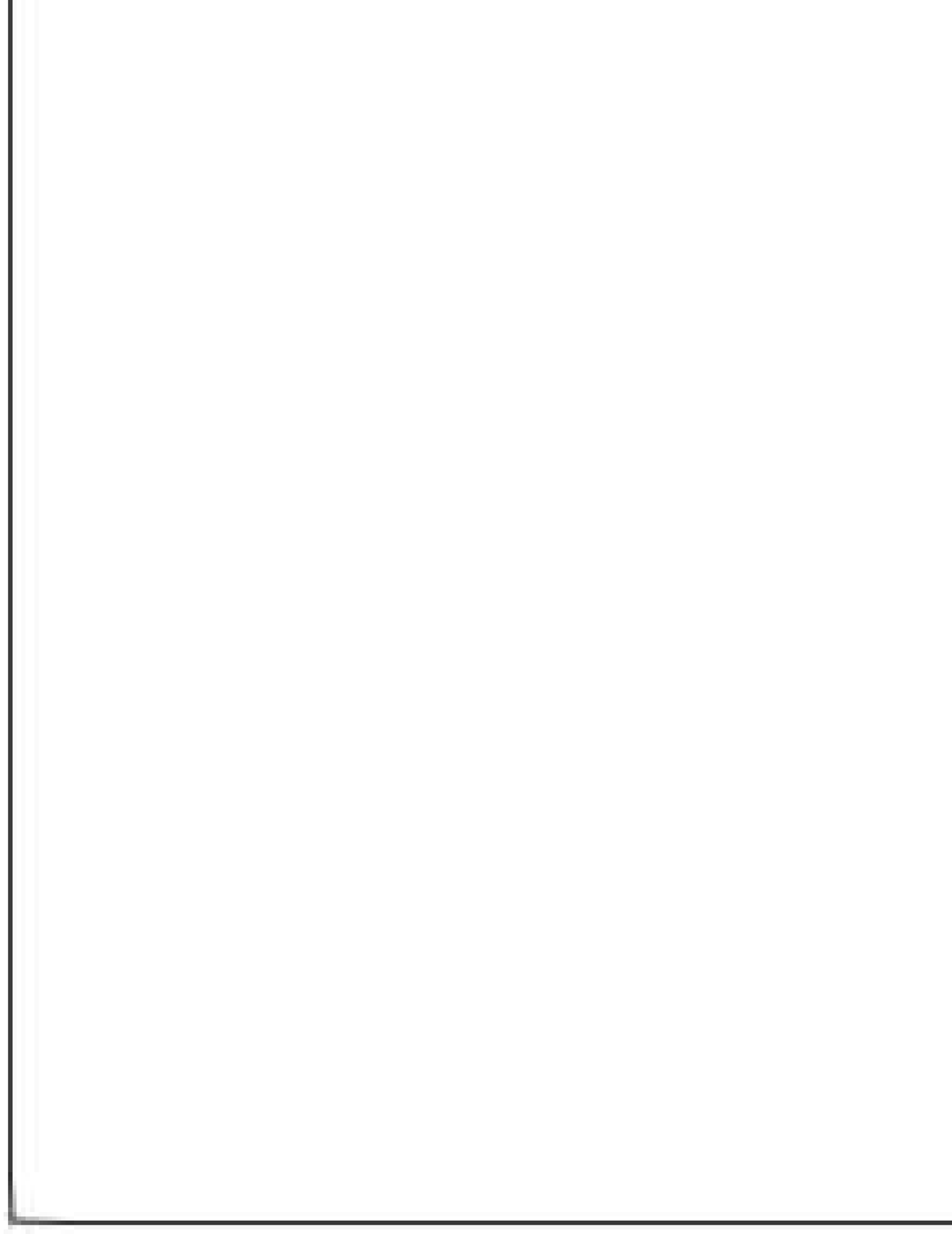} 
	\includegraphics[width=0.2\textwidth]{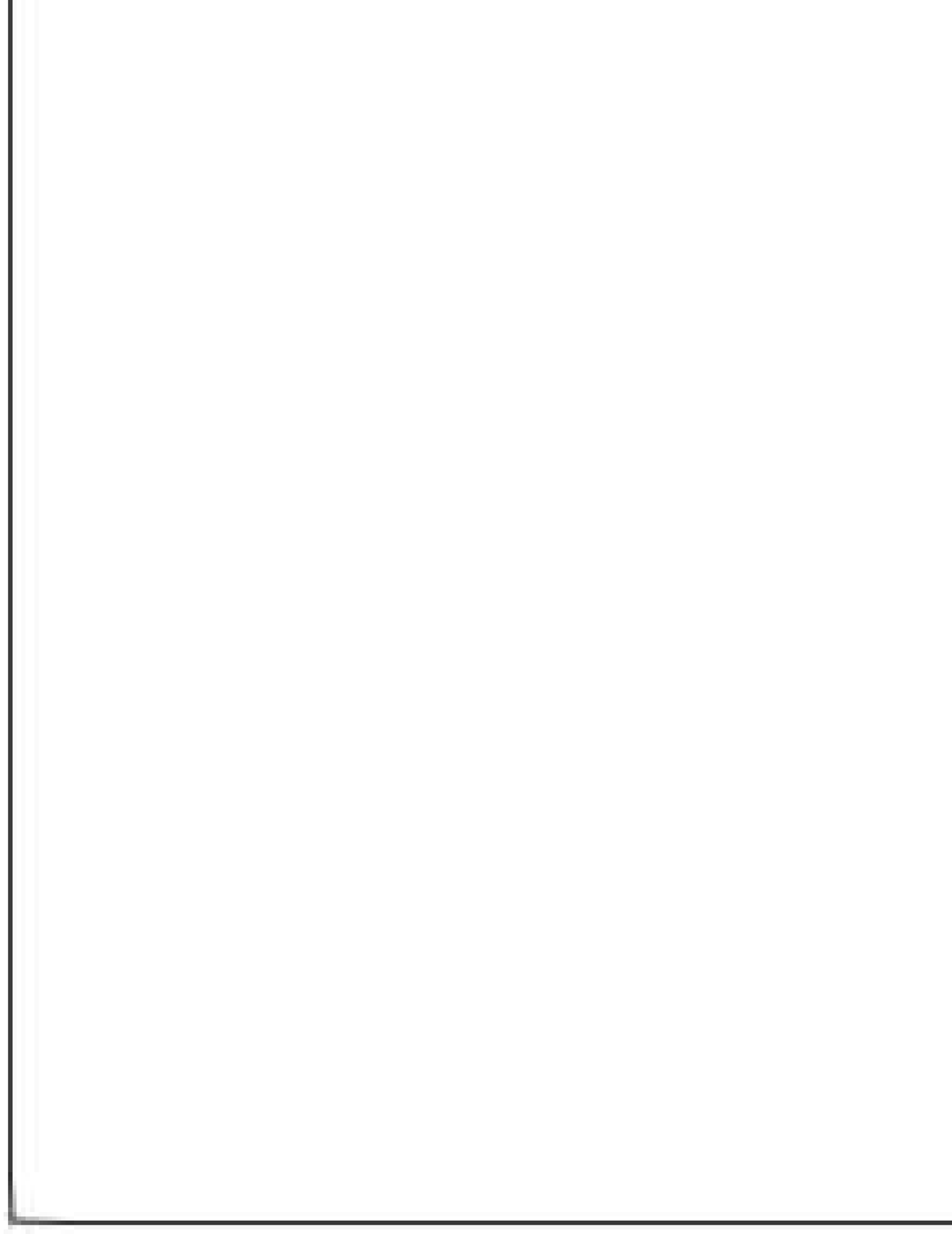} 
	\includegraphics[width=0.2\textwidth]{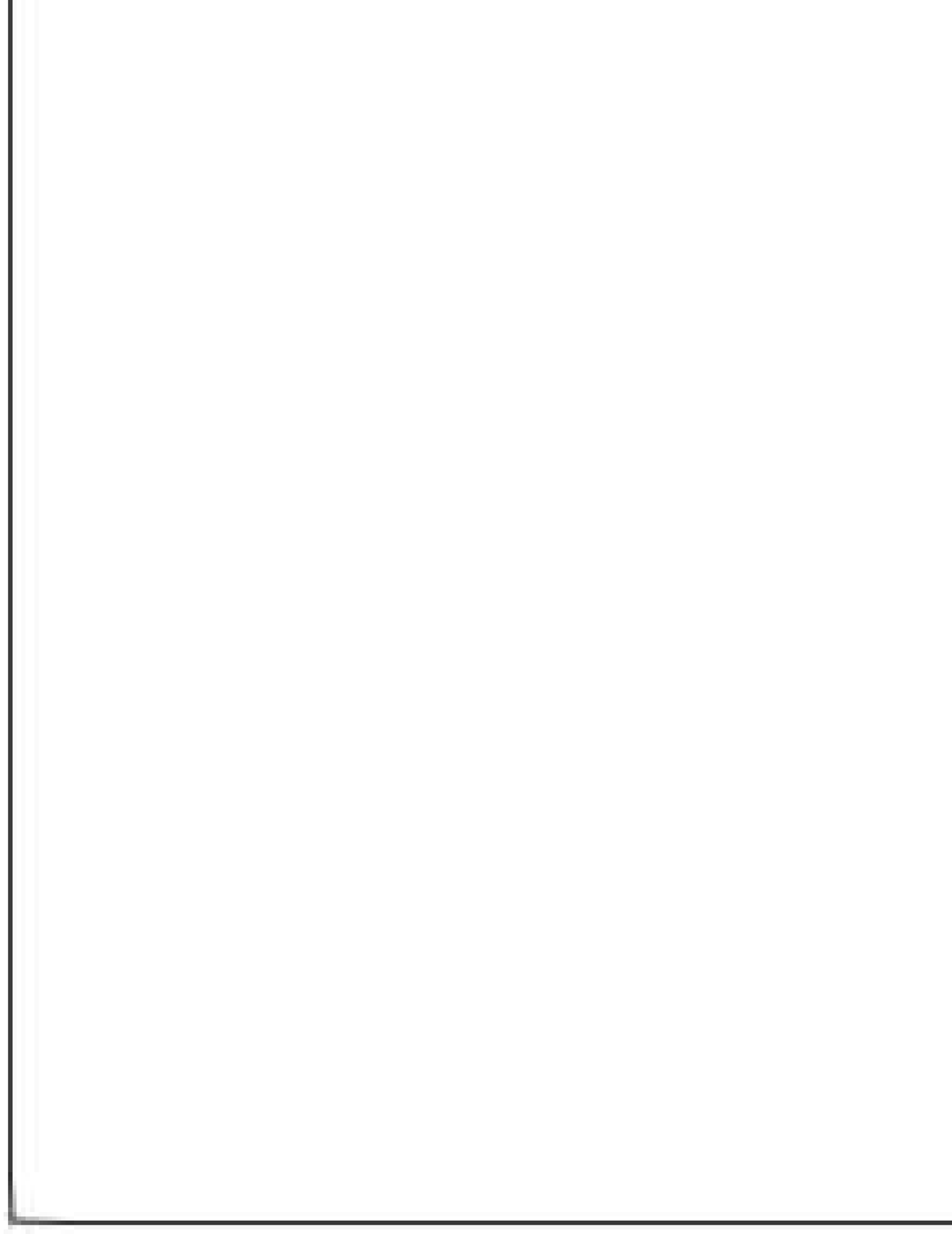}
	\put(-400,47){ (\textit{a})}
	\put(-382,47){\small $\mathit{Fo}=0.0$} 	
	\put(-286,47){\small $\mathit{Fo}=0.63$} 	 	
	\put(-190,47){\small $\mathit{Fo}=1.88$} 	 	
	\put(-93,47){\small $\mathit{Fo}=2.78$} 	
	\vspace{0.2cm}
	
	\includegraphics[width=0.2\textwidth]{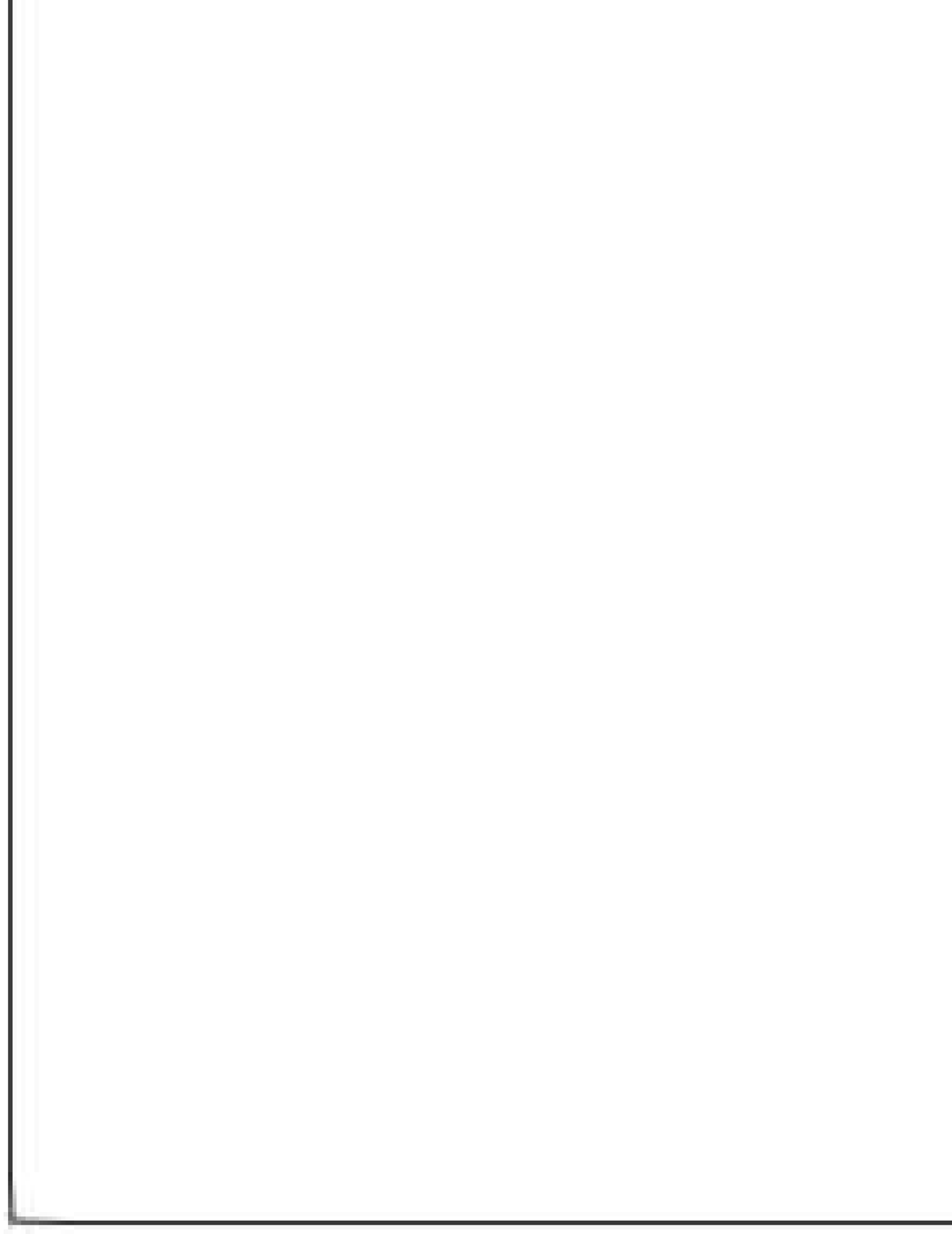} 
	\includegraphics[width=0.2\textwidth]{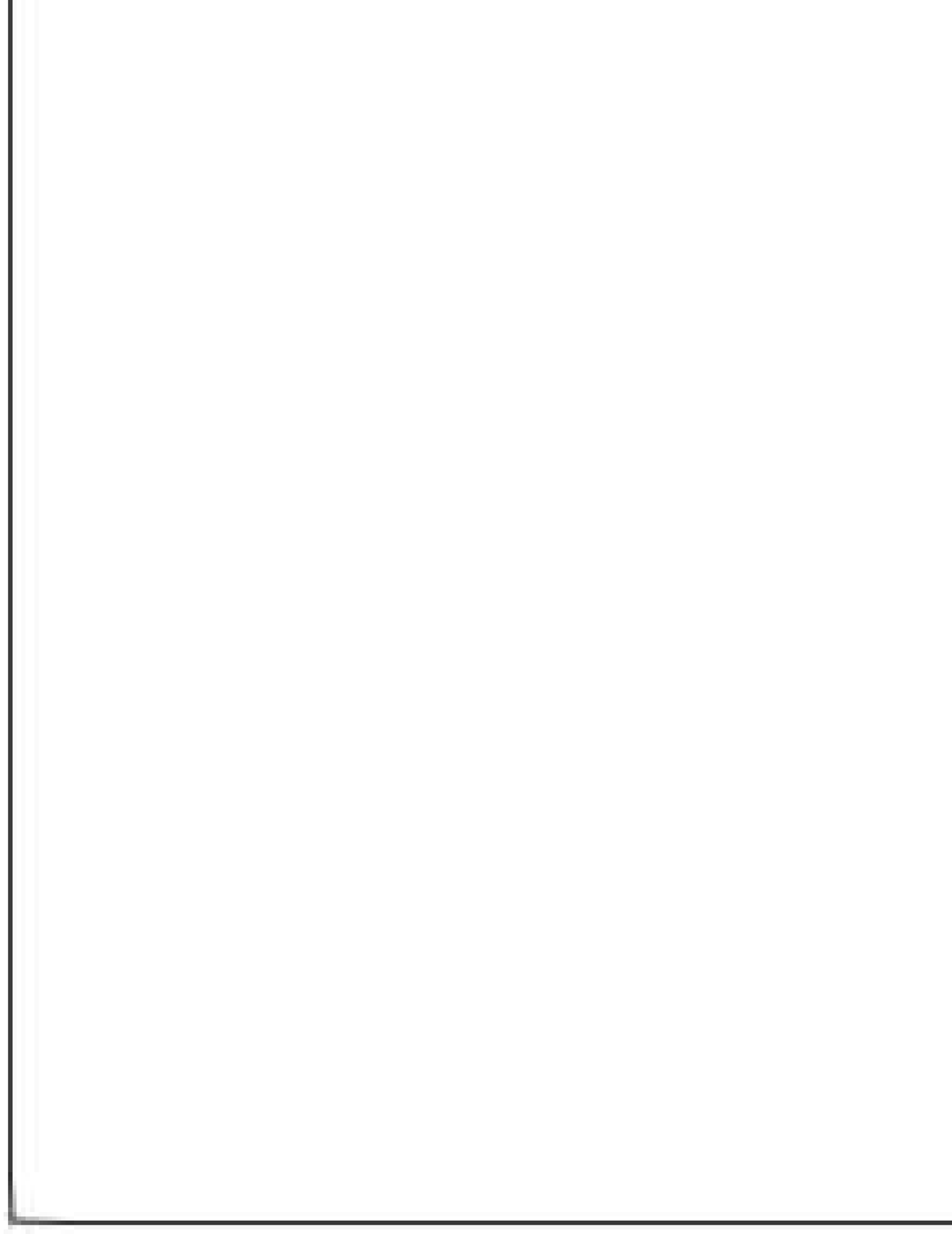} 
	\includegraphics[width=0.2\textwidth]{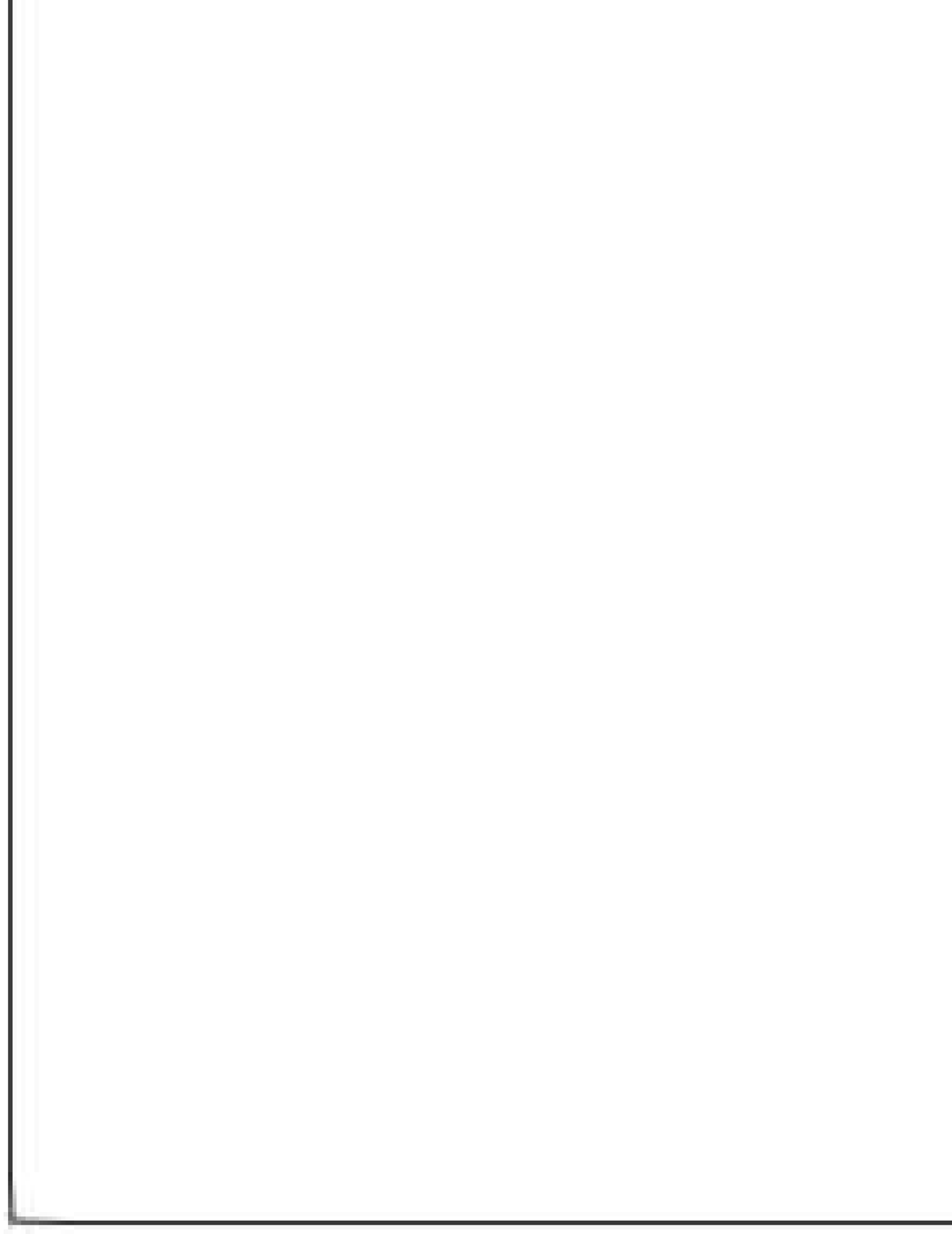} 
	\includegraphics[width=0.2\textwidth]{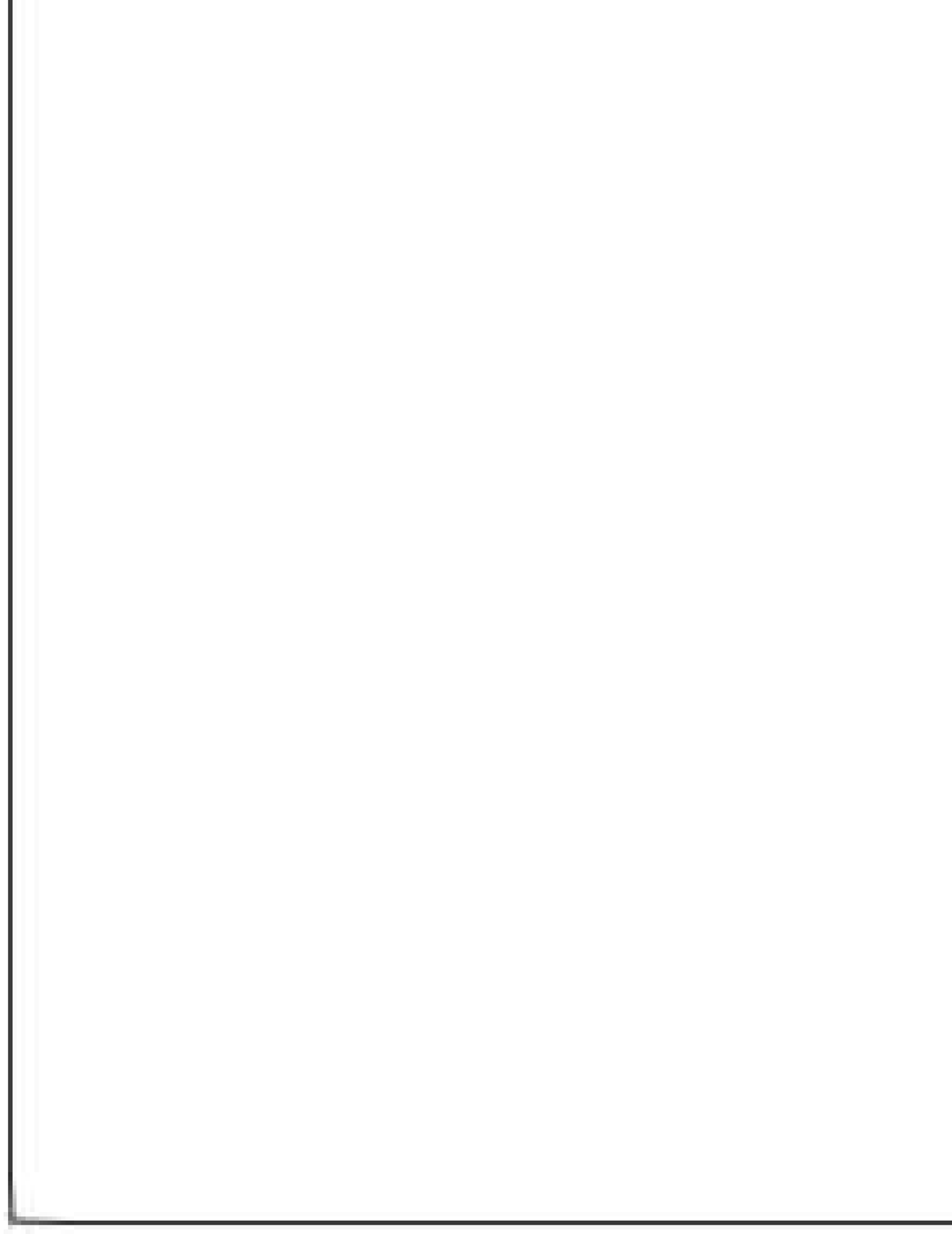}
	\put(-400,47){ (\textit{b})}
	\put(-382,47){\small $\mathit{Fo}=0.0$} 	
	\put(-286,47){\small $\mathit{Fo}=0.63$} 	 	
	\put(-190,47){\small $\mathit{Fo}=1.88$} 	 	
	\put(-93,47){\small $\mathit{Fo}=2.78$} 	
	\vspace{0.2cm}
	
	\includegraphics[width=0.2\textwidth]{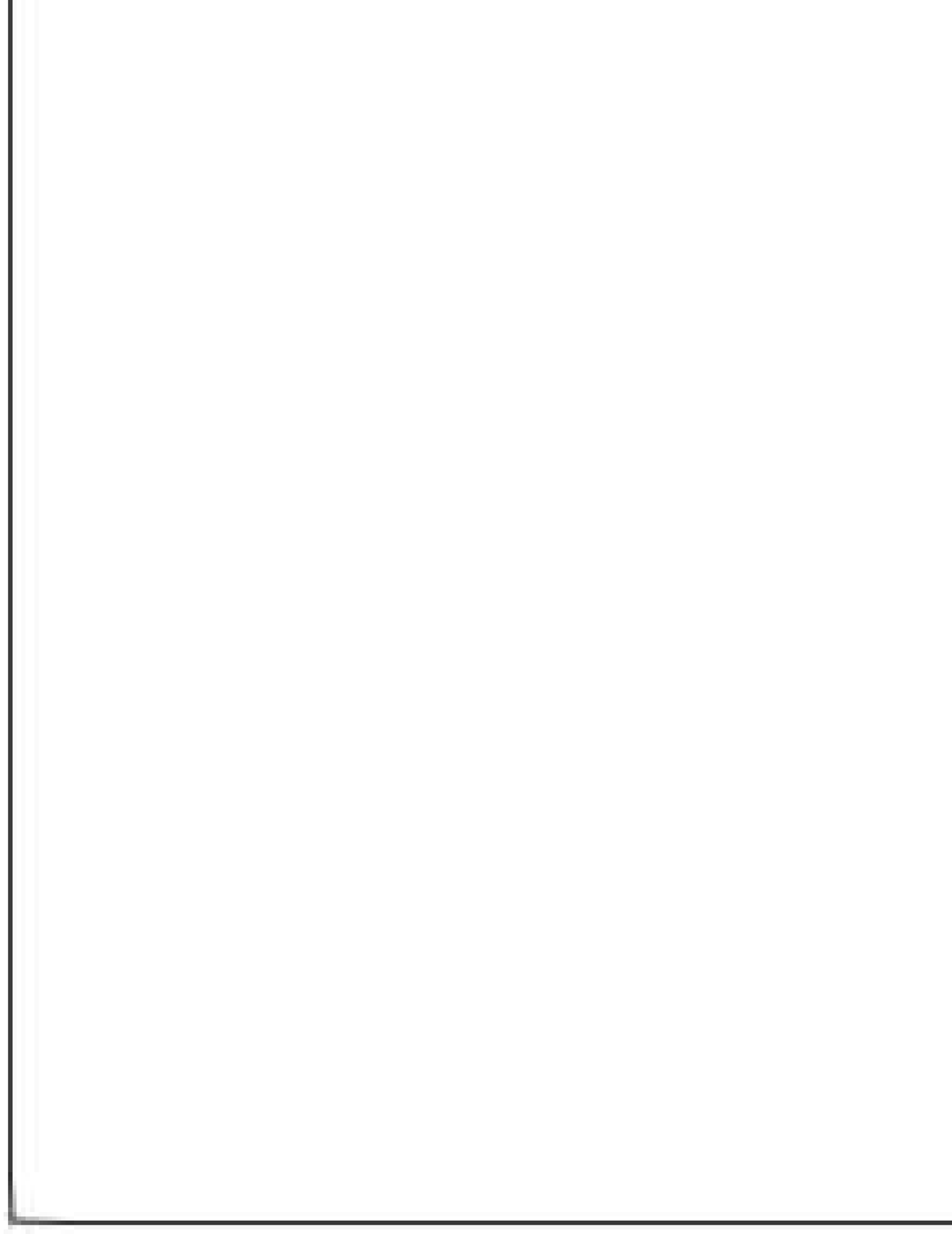} 
	\includegraphics[width=0.2\textwidth]{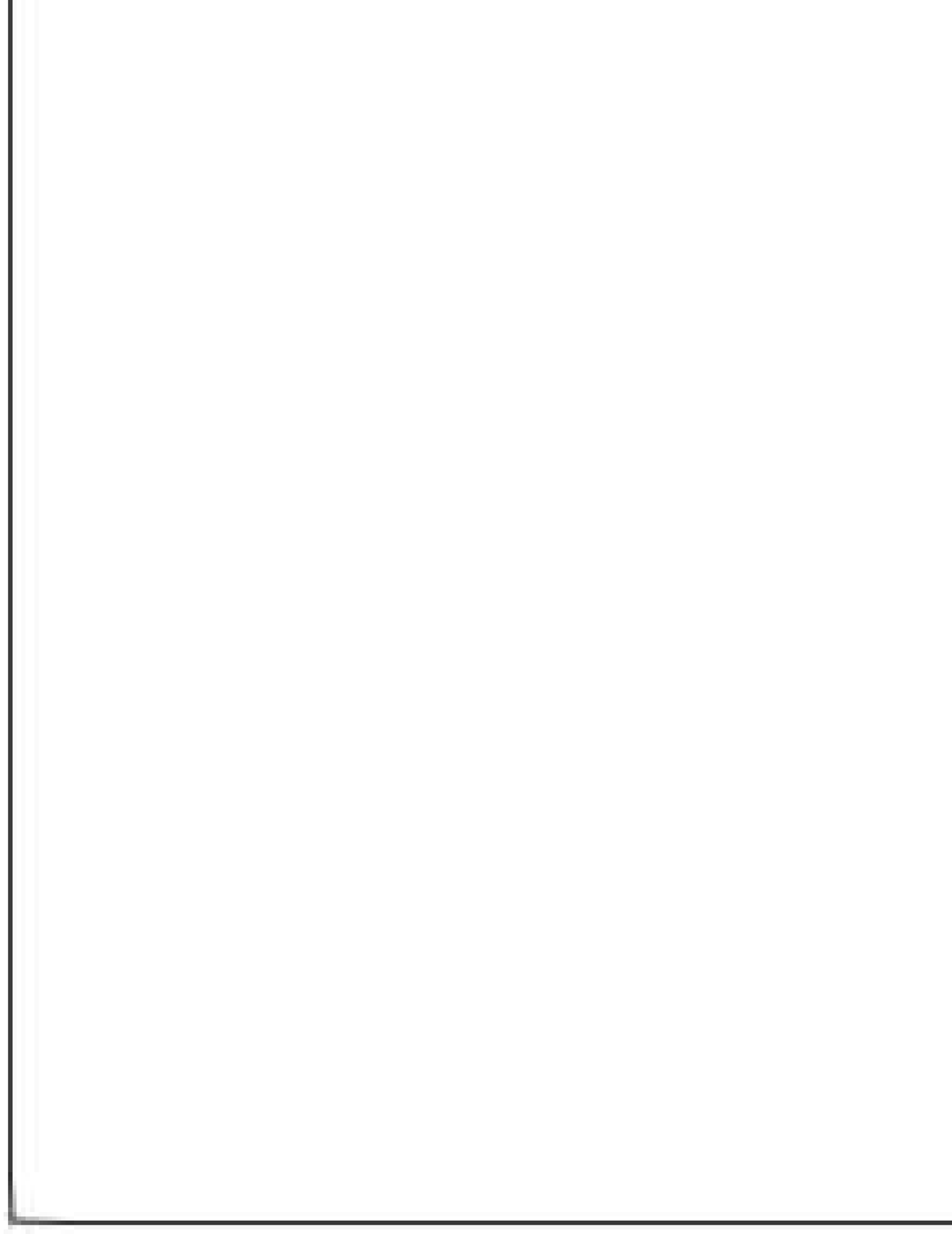} 
	\includegraphics[width=0.2\textwidth]{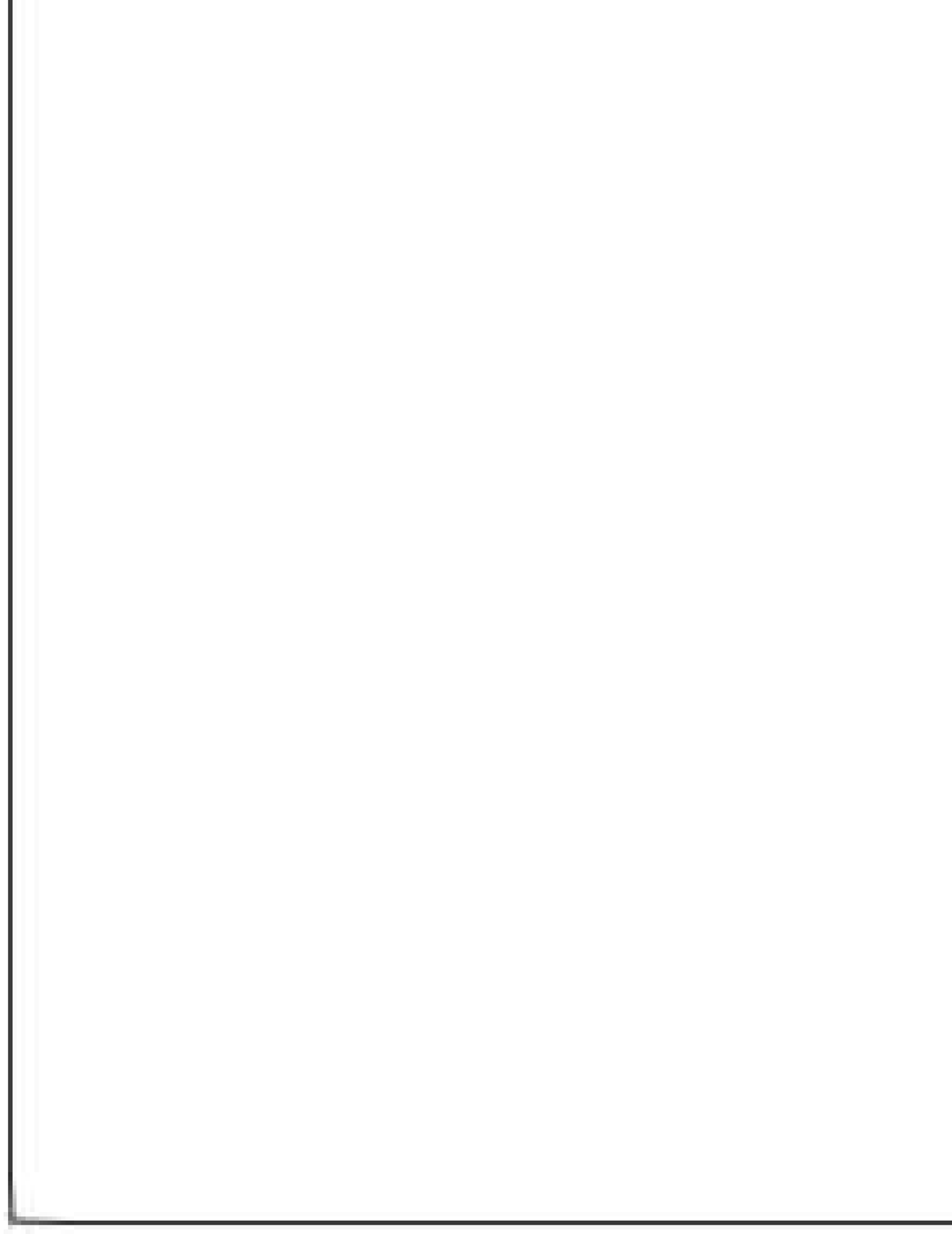} 
	\includegraphics[width=0.2\textwidth]{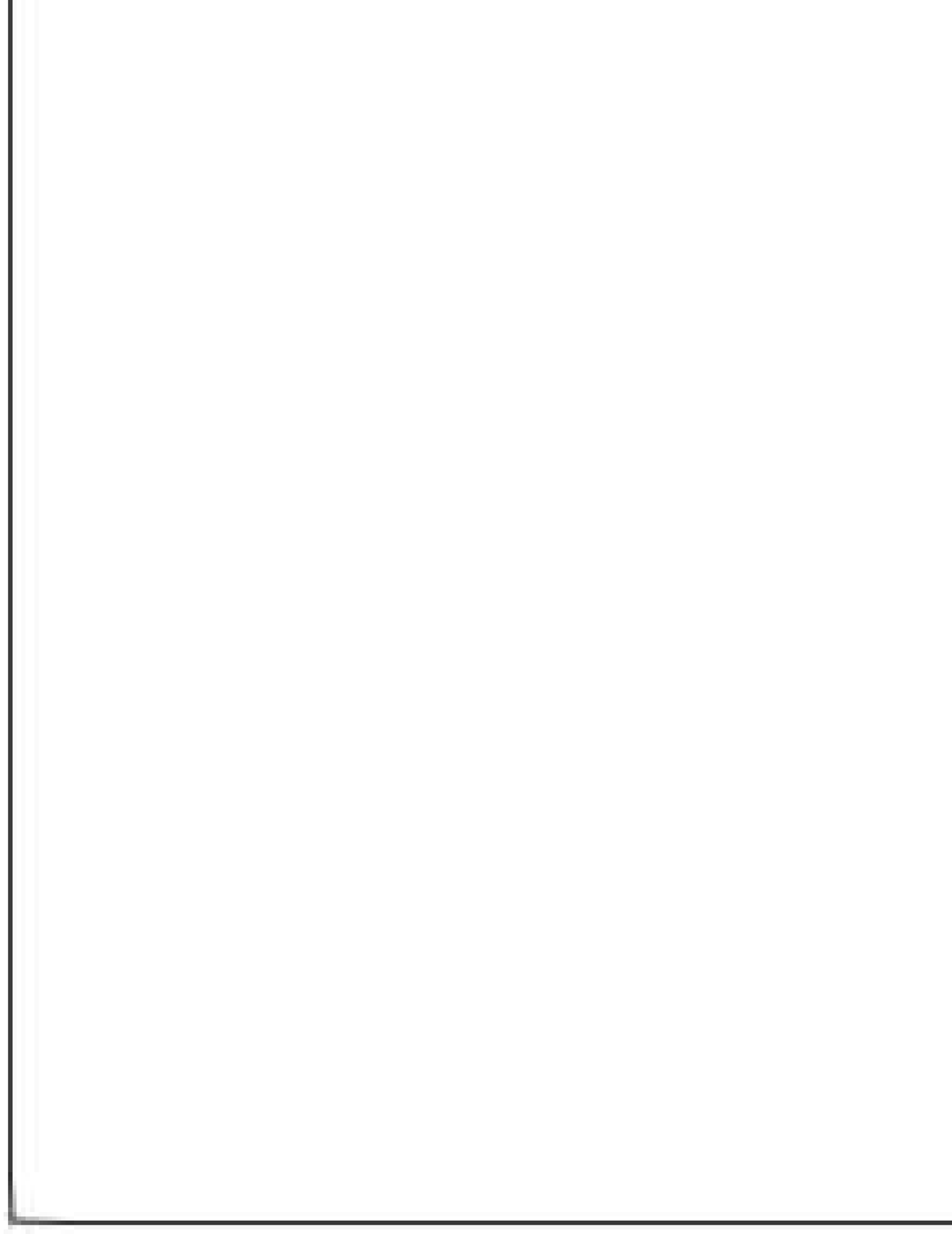}
	\put(-400,47){ (\textit{c})}
	\put(-382,47){\small $\mathit{Fo}=0.0$} 	
	\put(-286,47){\small $\mathit{Fo}=0.63$} 	 	
	\put(-190,47){\small $\mathit{Fo}=1.88$} 	 	
	\put(-93,47){\small $\mathit{Fo}=2.78$} 		 	
	\vspace{0.2cm}
	
	\includegraphics[width=0.2\textwidth]{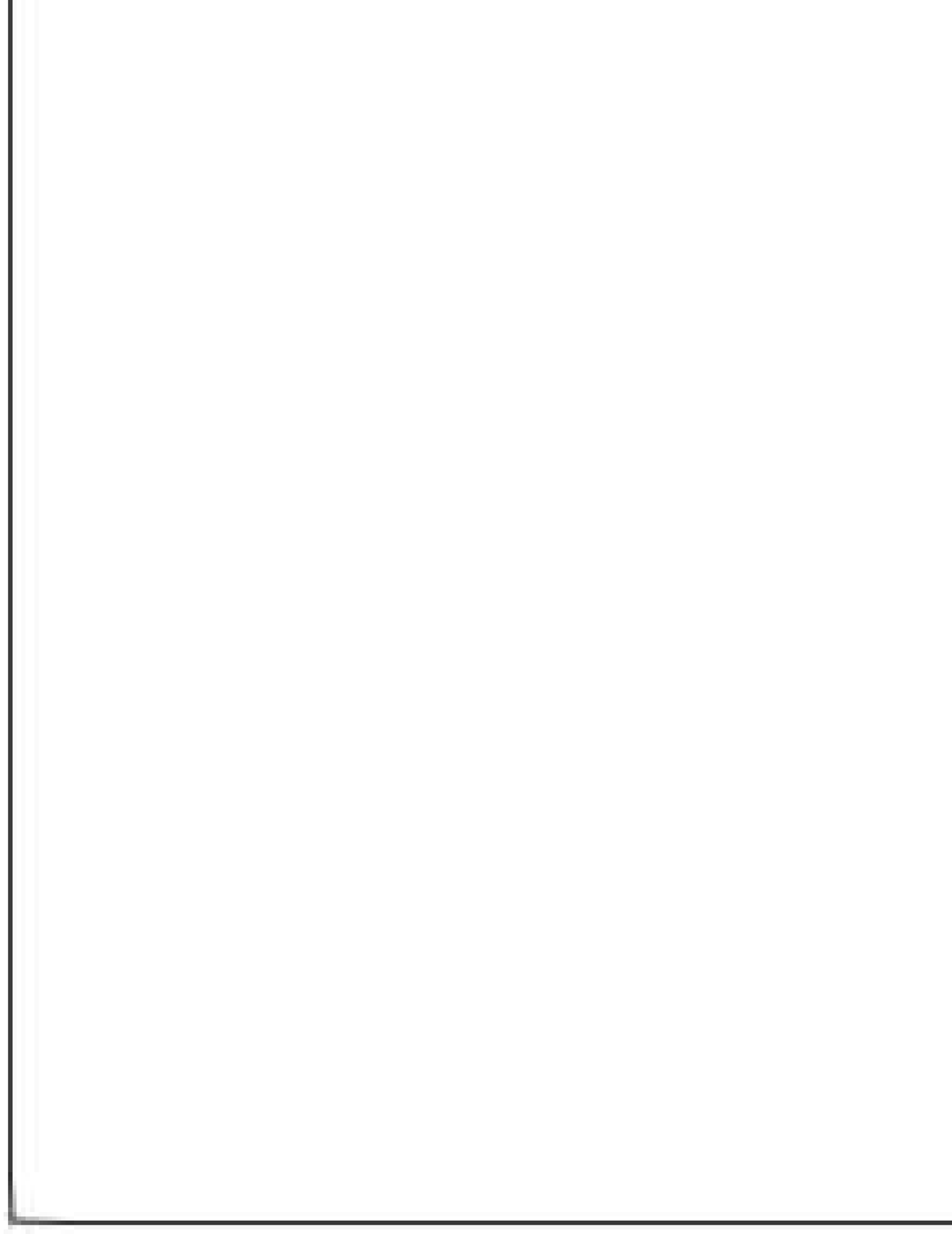} 
	\includegraphics[width=0.2\textwidth]{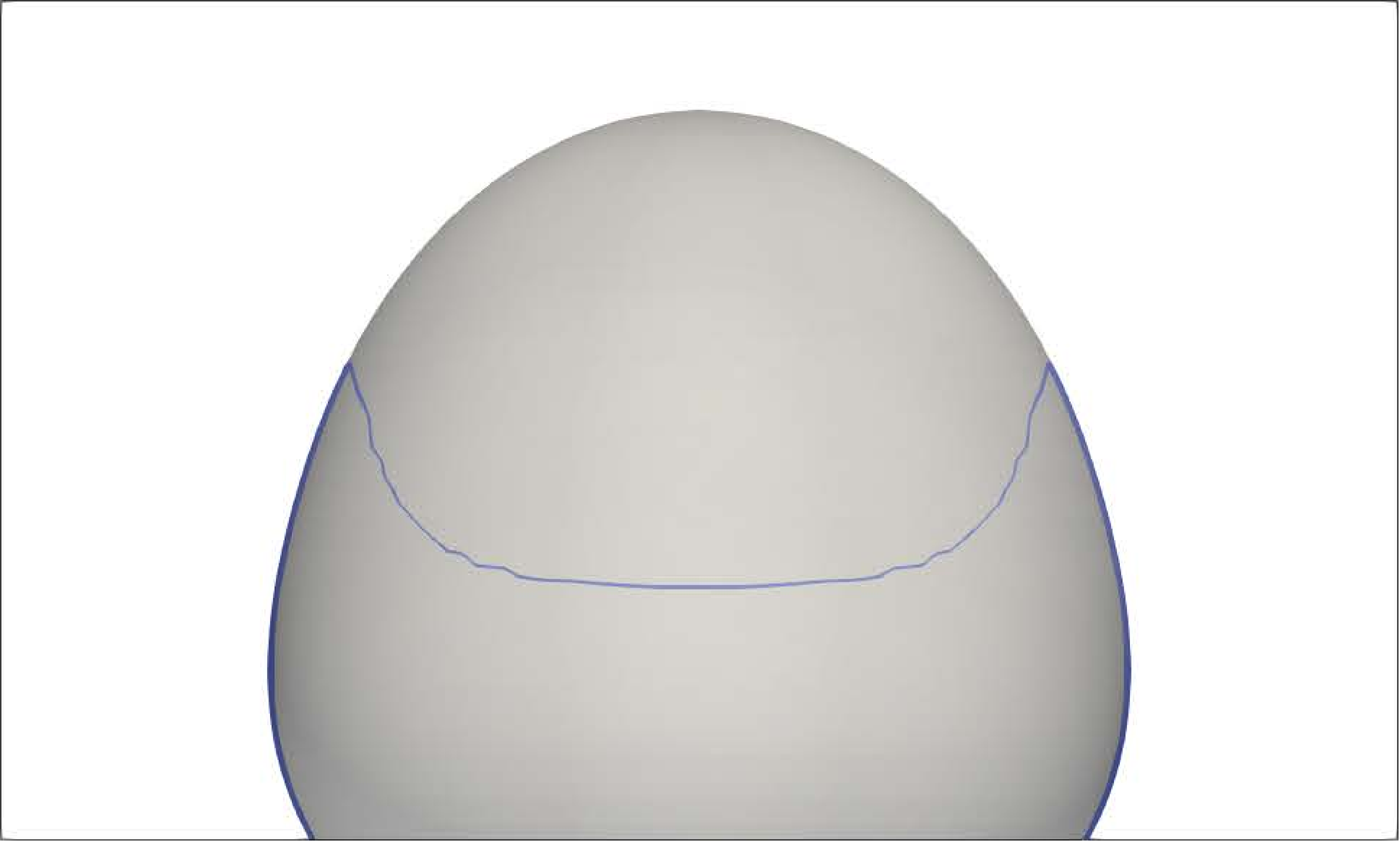} 
	\includegraphics[width=0.2\textwidth]{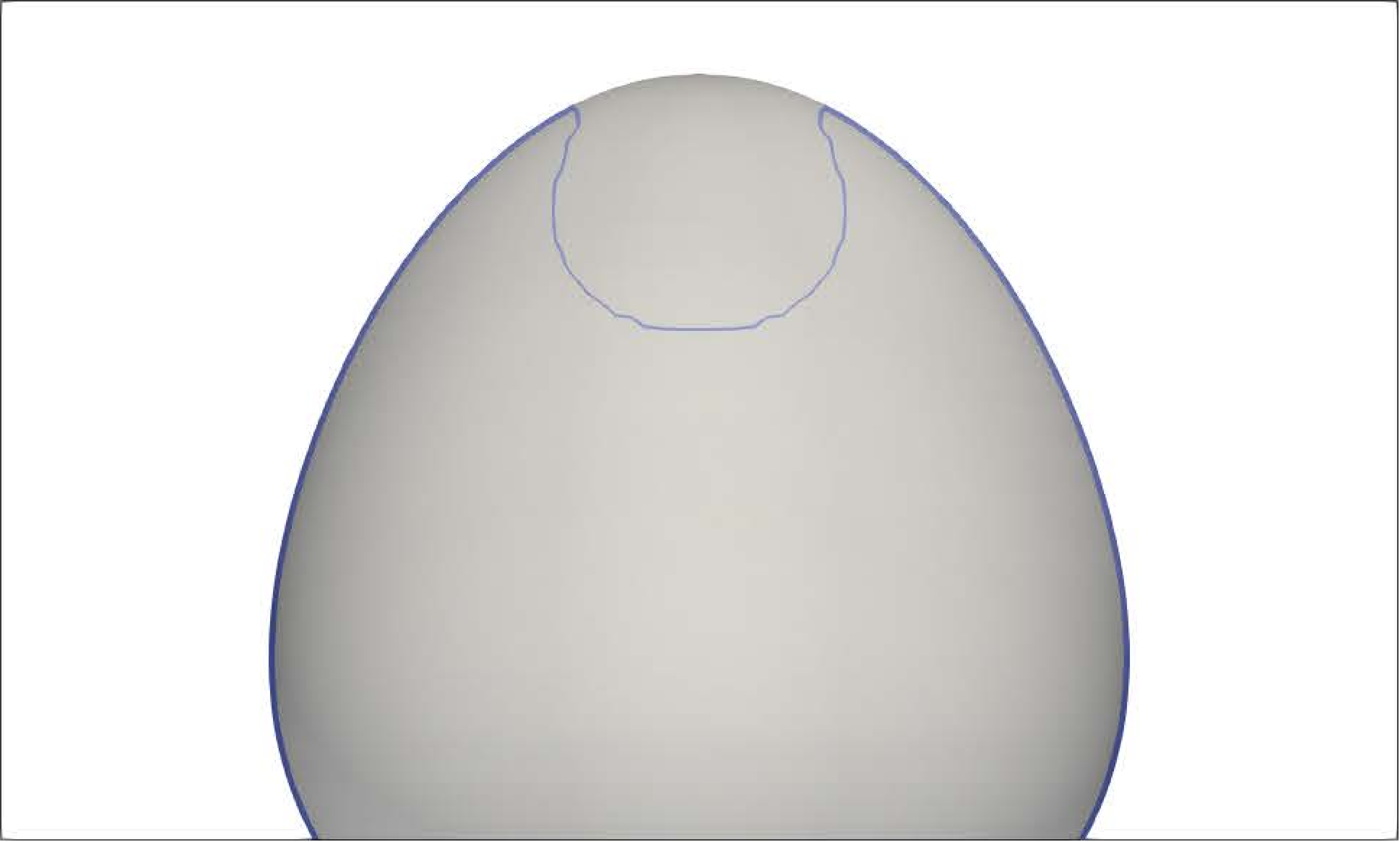} 
	\includegraphics[width=0.2\textwidth]{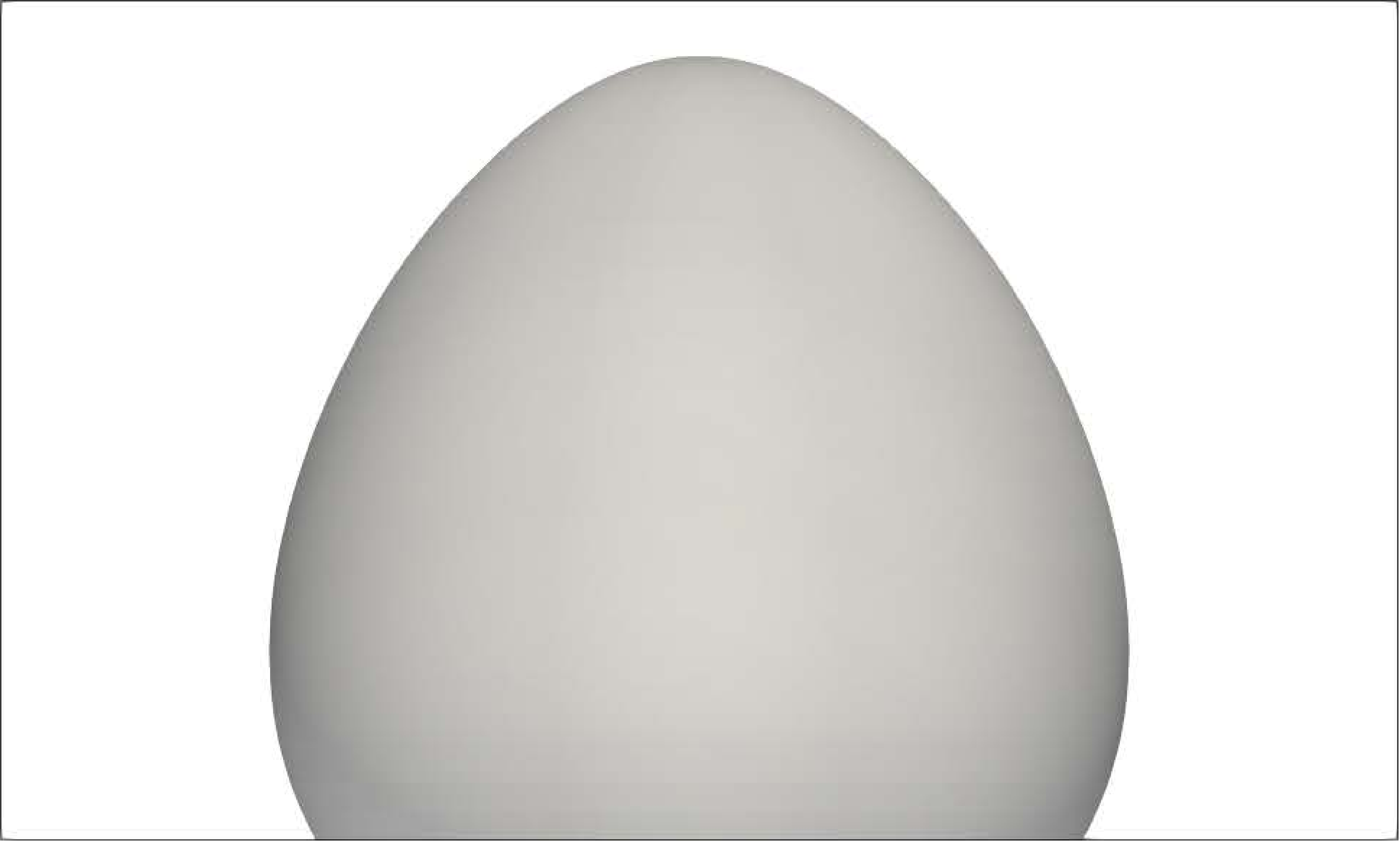}
	\put(-400,47){ (\textit{d})}
	\put(-382,47){\small $\mathit{Fo}=0.0$} 	
	\put(-286,47){\small $\mathit{Fo}=0.63$} 	 	
	\put(-190,47){\small $\mathit{Fo}=1.88$} 	 	
	\put(-93,47){\small $\mathit{Fo}=2.78$} 		 	
	\vspace{0.2cm}

	\caption{Evolutions of freezing front (solid line) and droplet profile for $Ca_E=0.0$ (a), $PR^+:R=2.0, S=0.5, Ca_E=1.0$ (b), $OB^-:R=2.0, S=5.0, Ca_E=1.0$ (c), and $PR^-:R=5.0, S=9.5, Ca_E=1.0$ (d). Other parameters are set as $\mathit{Ste}=0.2, \rho_s/\rho_l=0.9, \alpha_l/\alpha_g=0.2, \theta=120^{\circ}$. }
	\label{fig7}
\end{figure}

\begin{figure}[H]
	\centering
	\includegraphics[width=0.2\textwidth]{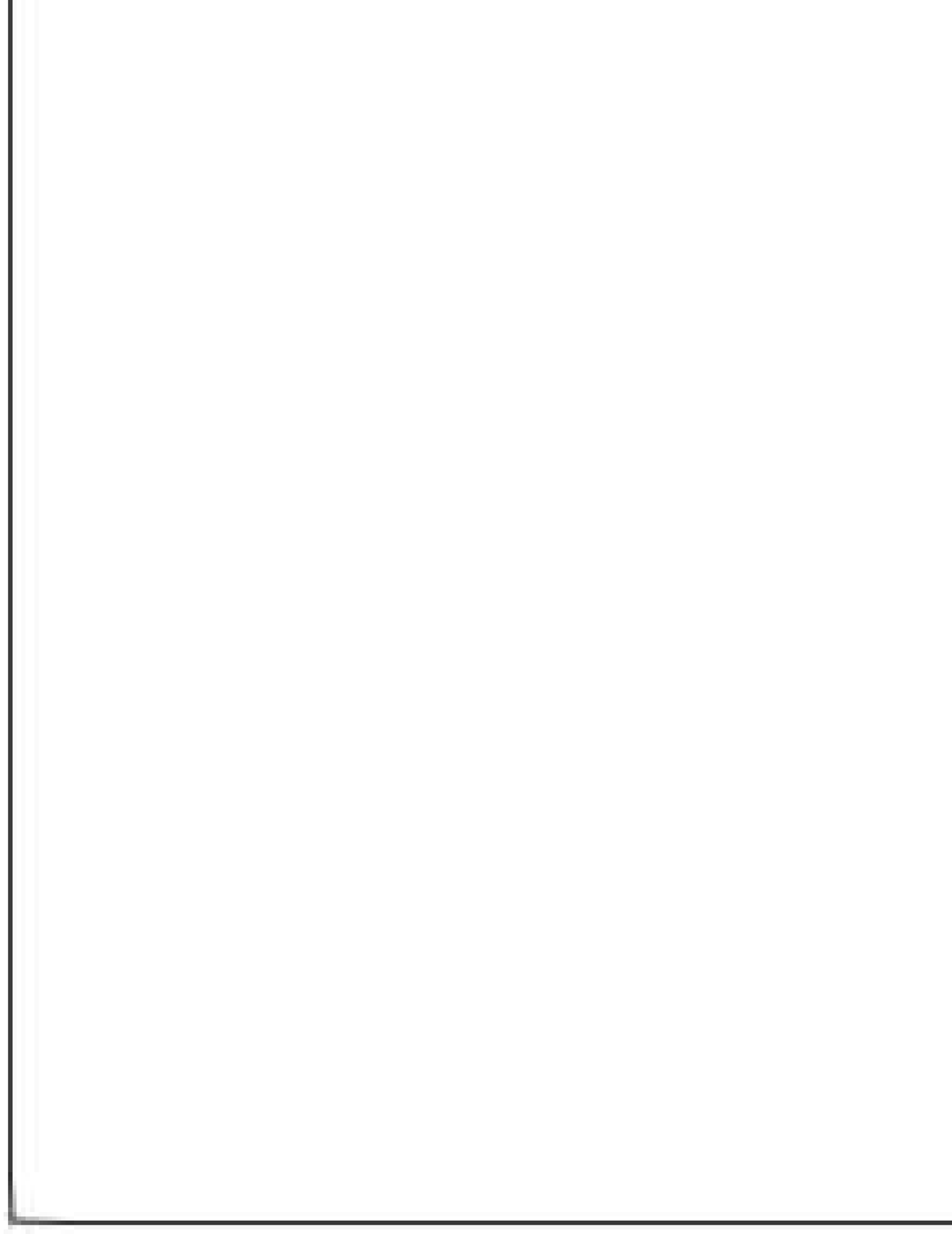} 
	\includegraphics[width=0.2\textwidth]{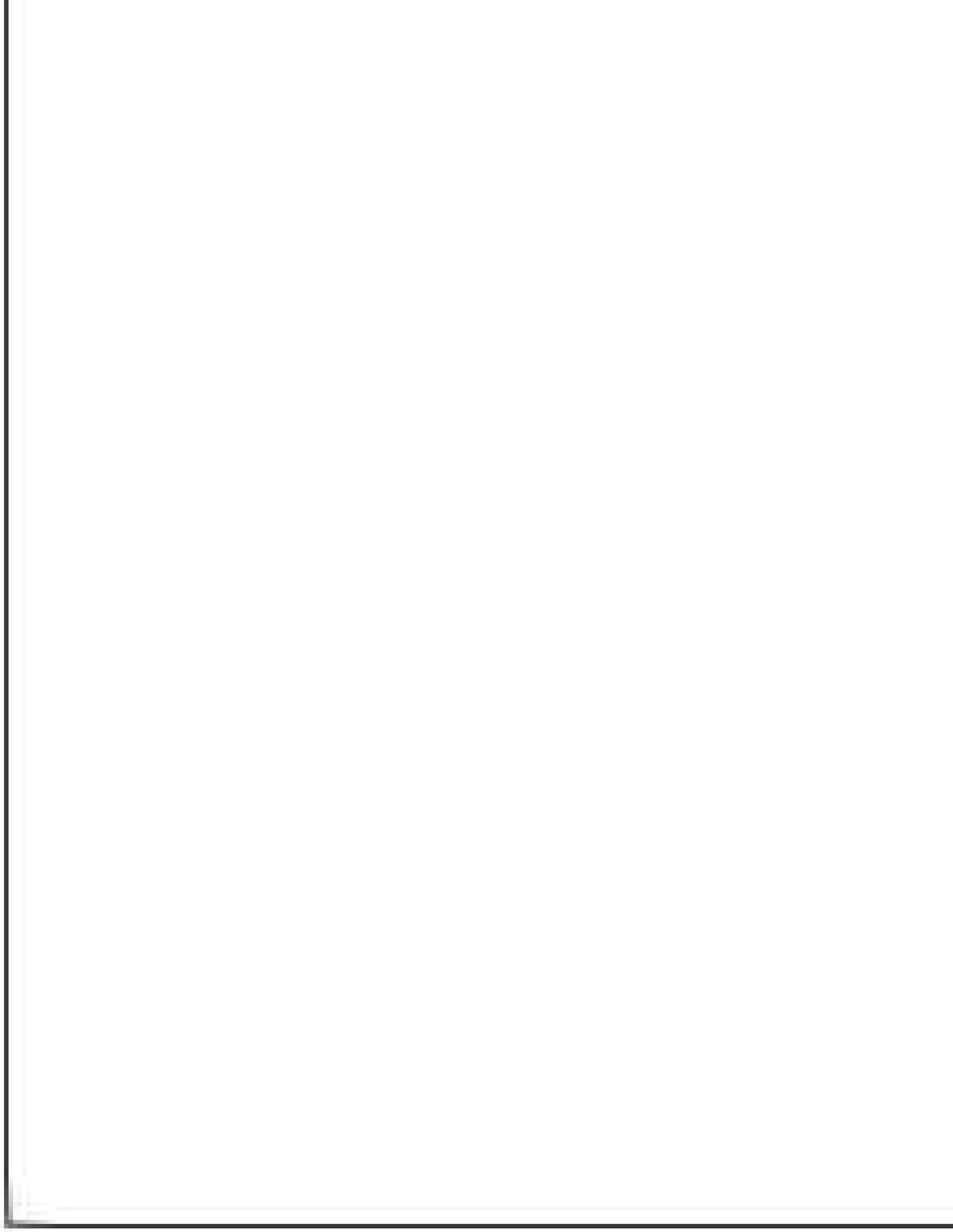} 
	\includegraphics[width=0.2\textwidth]{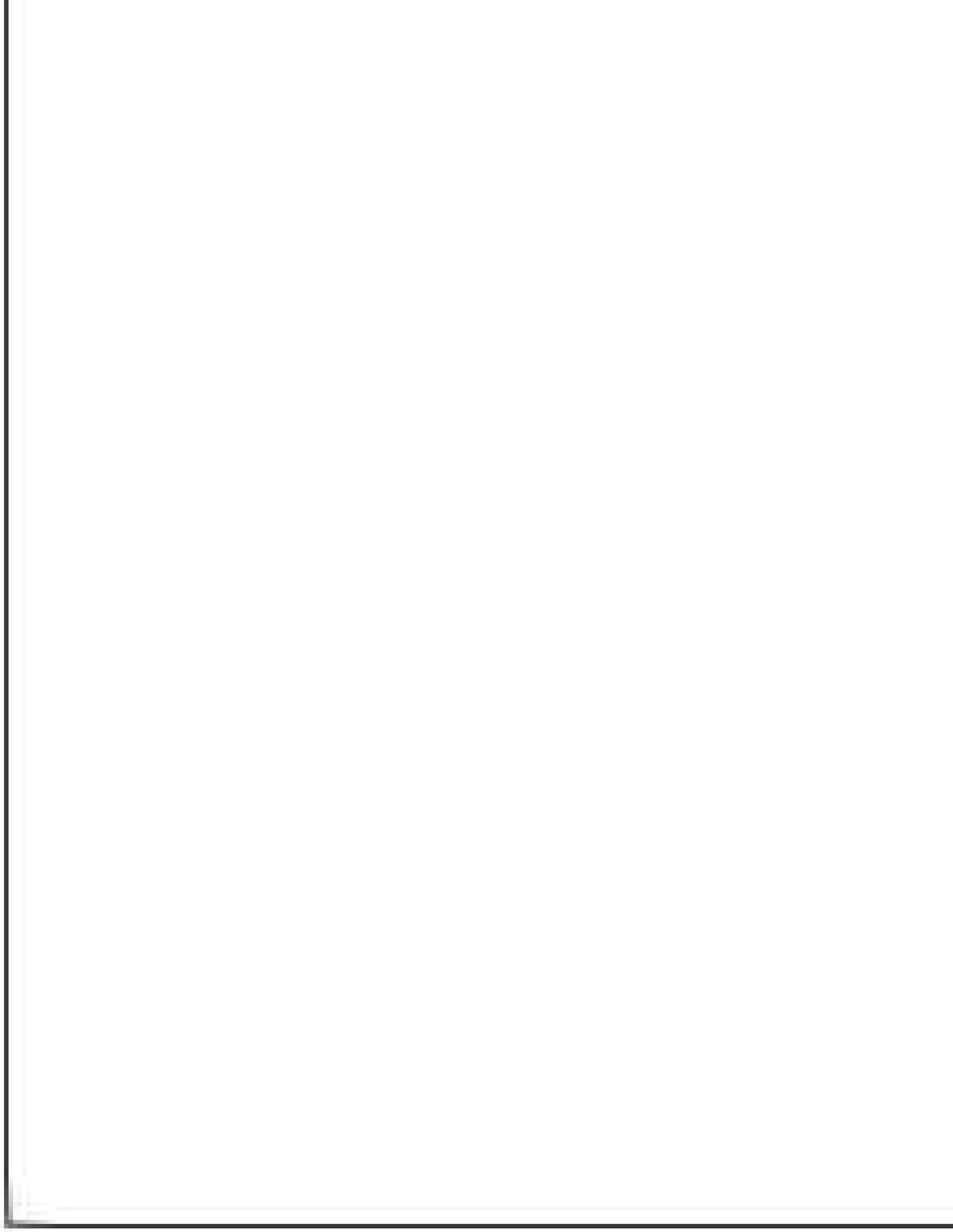} 
	\includegraphics[width=0.2\textwidth]{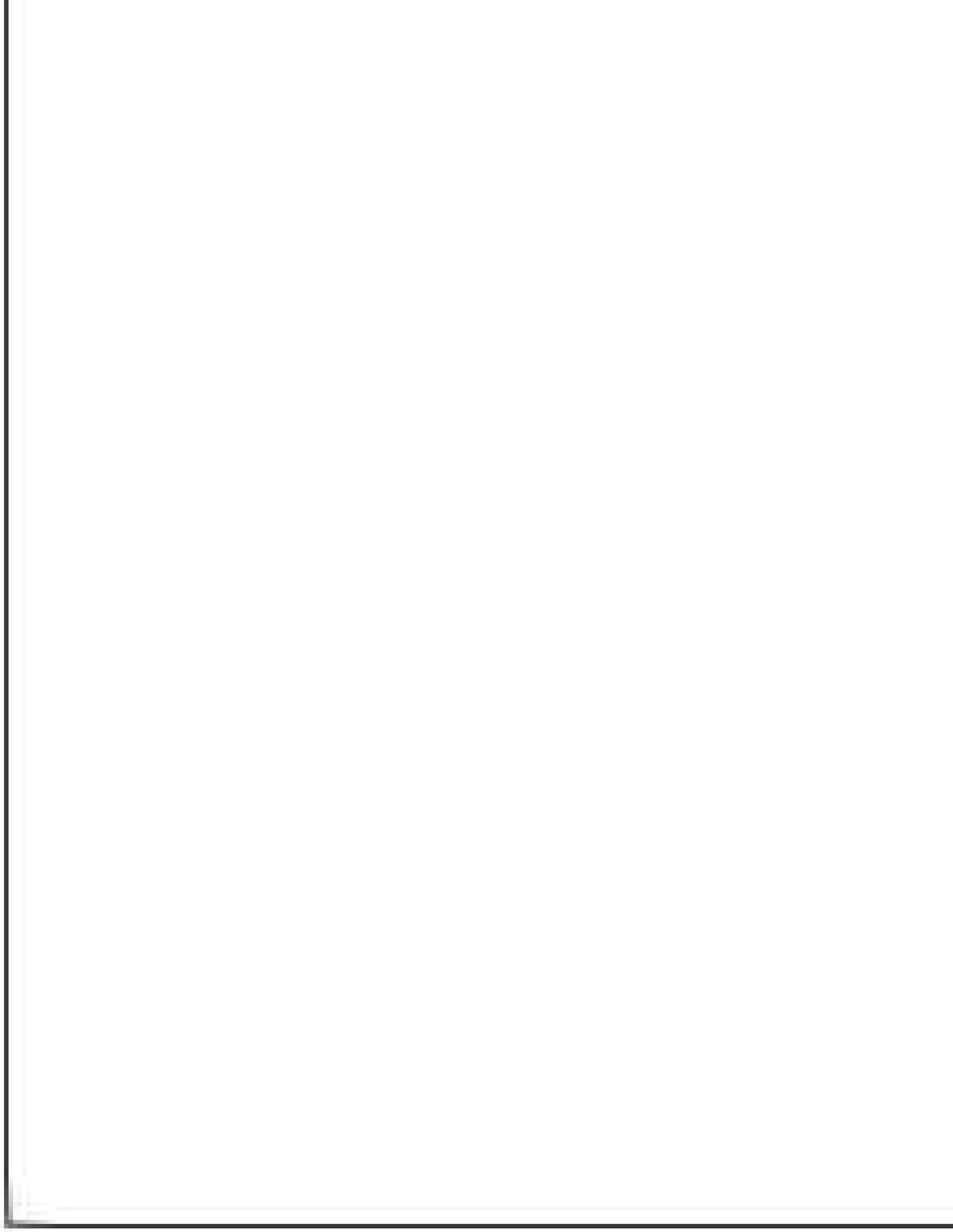}
	\put(-400,47){ (\textit{a})}
	\put(-382,47){\small $\mathit{Fo}=0.0$} 	
	\put(-286,47){\small $\mathit{Fo}=0.63$} 	 	
	\put(-190,47){\small $\mathit{Fo}=1.25$} 	 	
	\put(-93,47){\small $\mathit{Fo}=1.97$} 	
	\vspace{0.2cm}
	
	\includegraphics[width=0.2\textwidth]{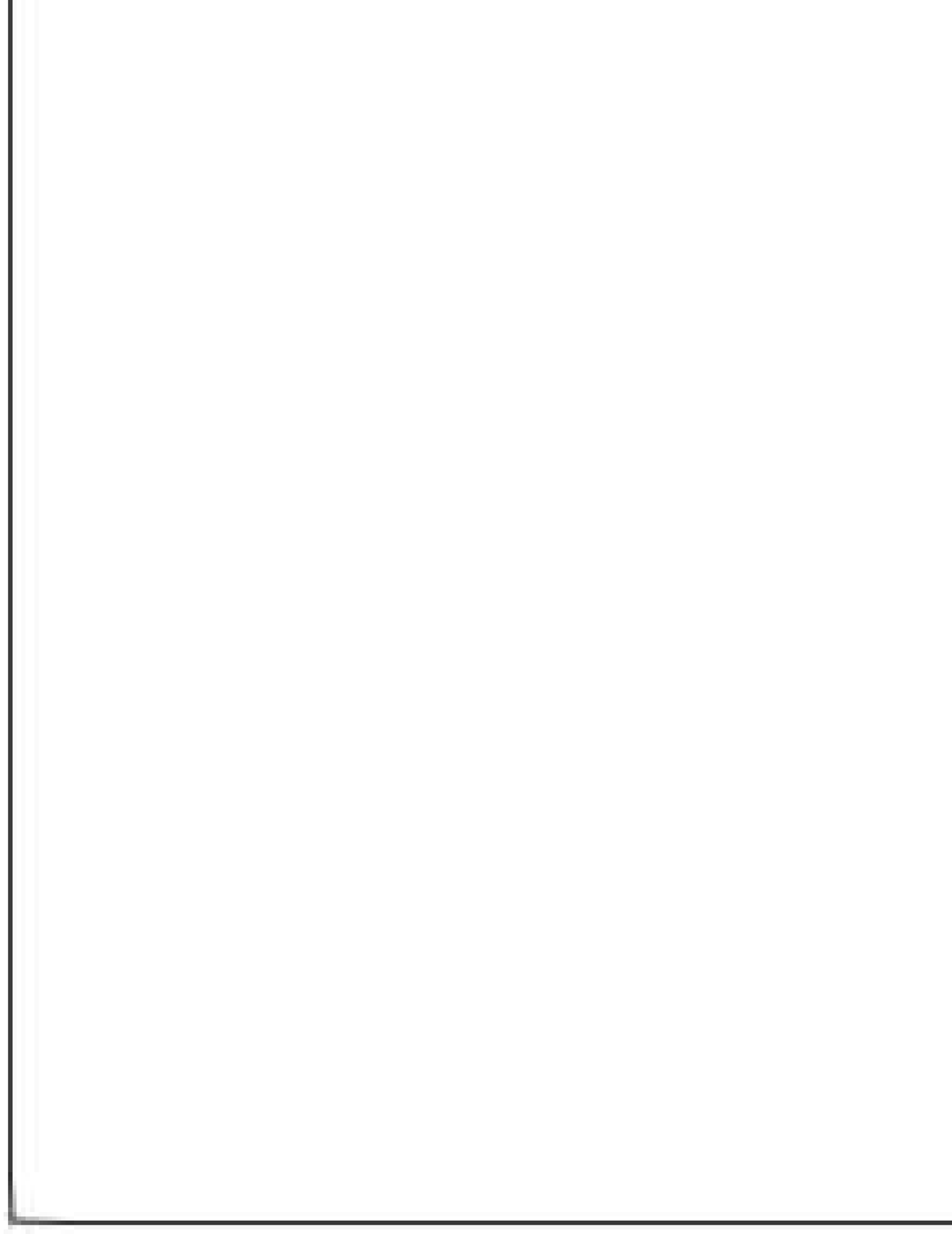} 
	\includegraphics[width=0.2\textwidth]{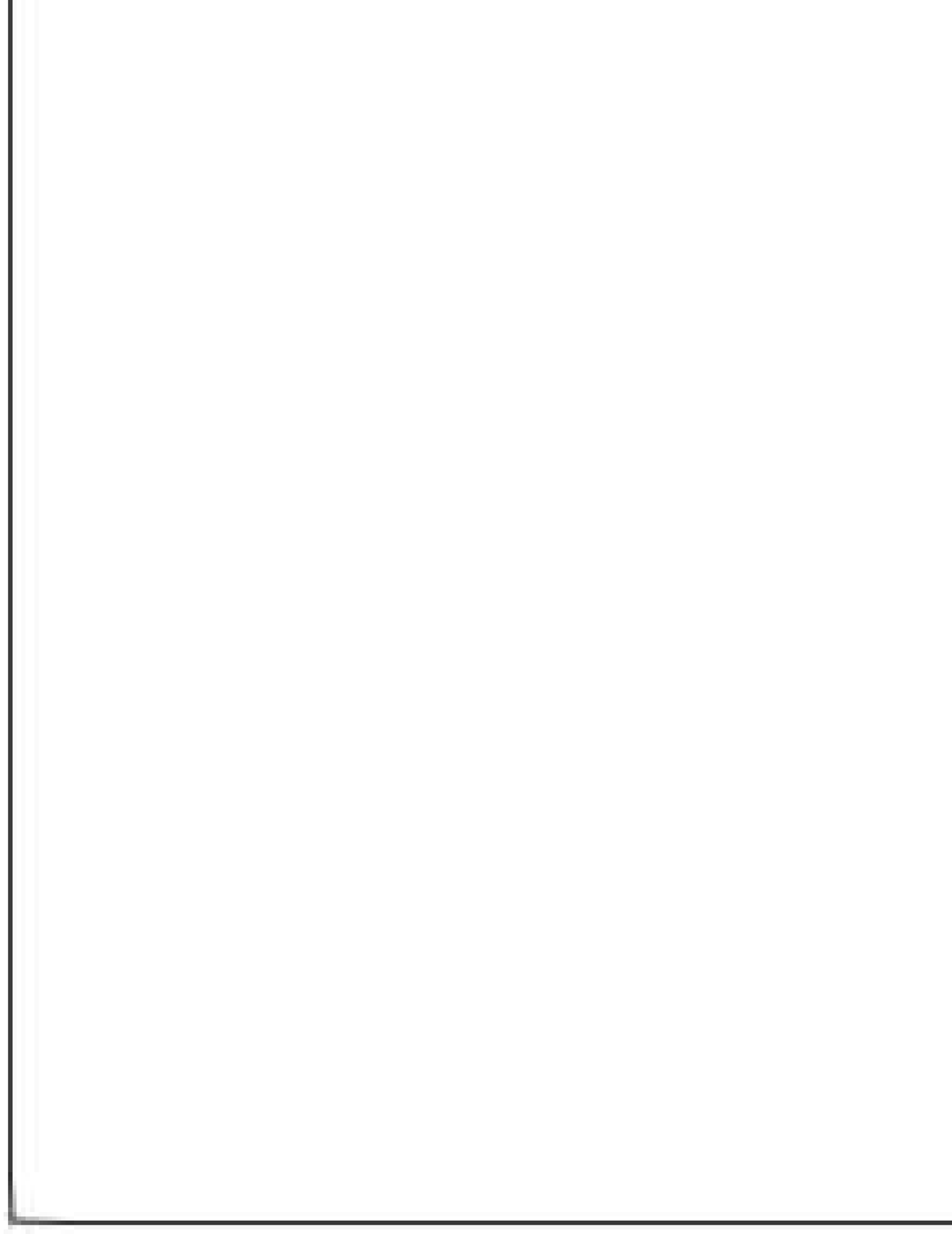} 
	\includegraphics[width=0.2\textwidth]{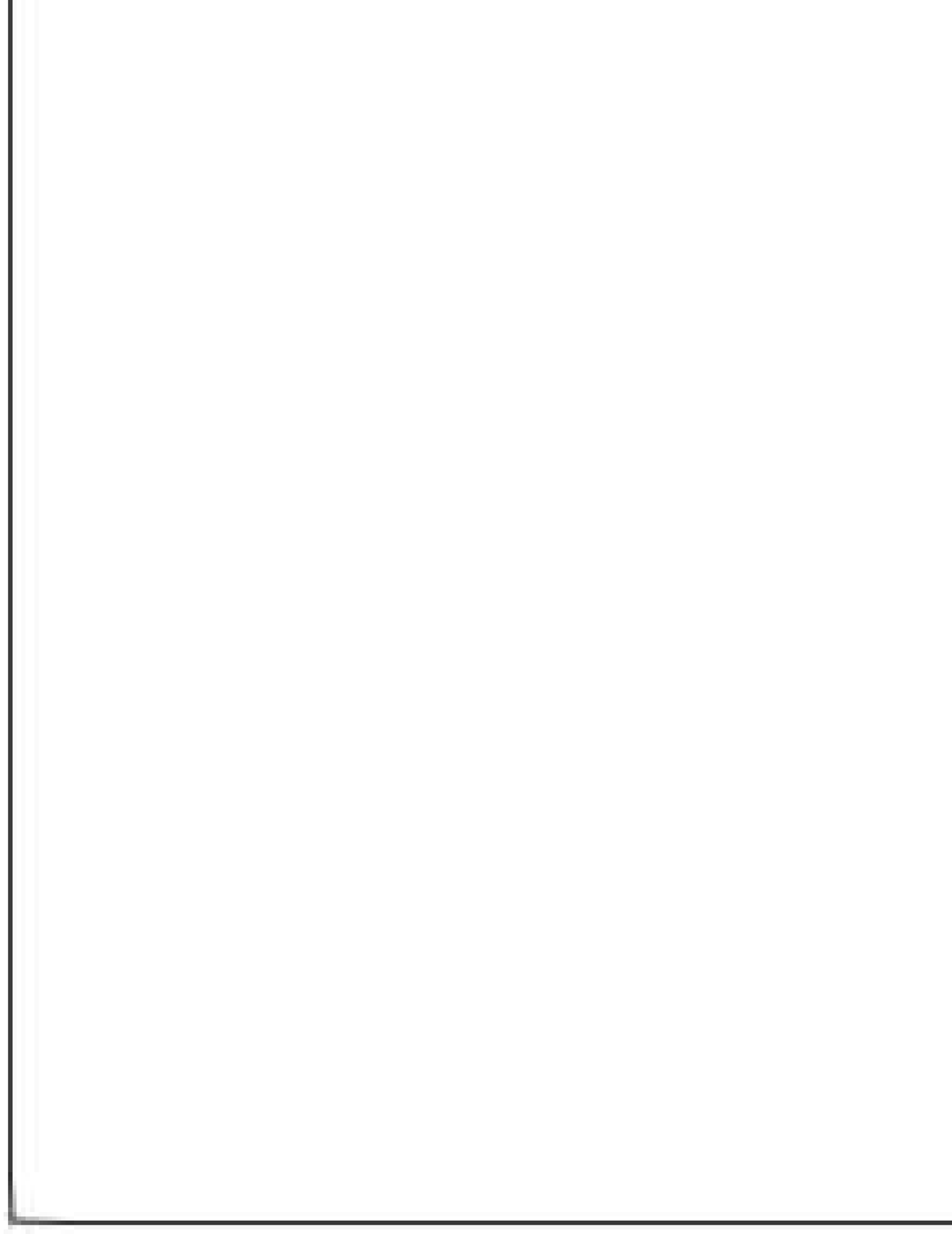} 
	\includegraphics[width=0.2\textwidth]{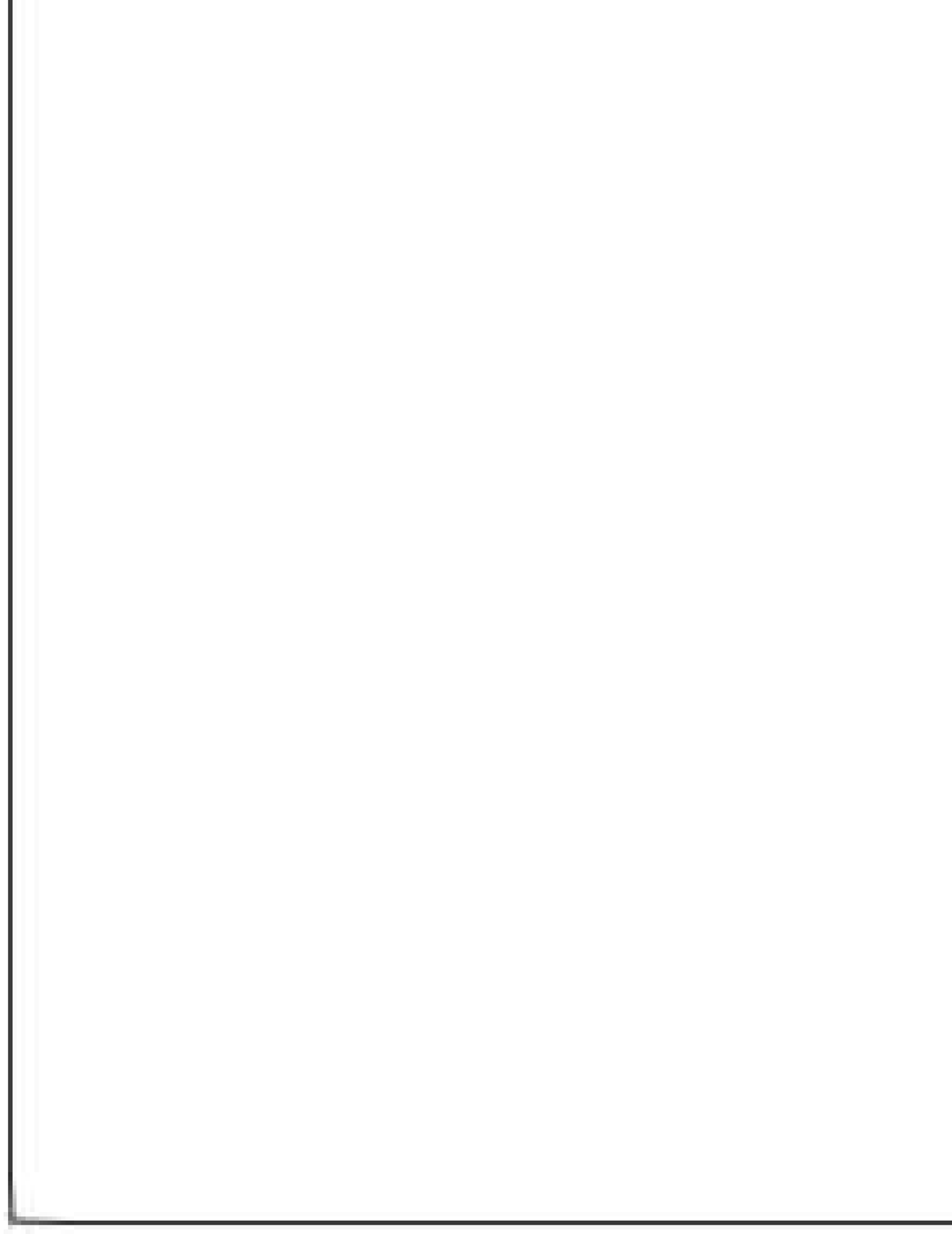}
	\put(-400,47){ (\textit{b})}
	\put(-382,47){\small $\mathit{Fo}=0.0$} 	
	\put(-286,47){\small $\mathit{Fo}=0.63$} 	 	
	\put(-190,47){\small $\mathit{Fo}=1.25$} 	 	
	\put(-93,47){\small $\mathit{Fo}=1.97$} 	
	\vspace{0.2cm}
	
	\includegraphics[width=0.2\textwidth]{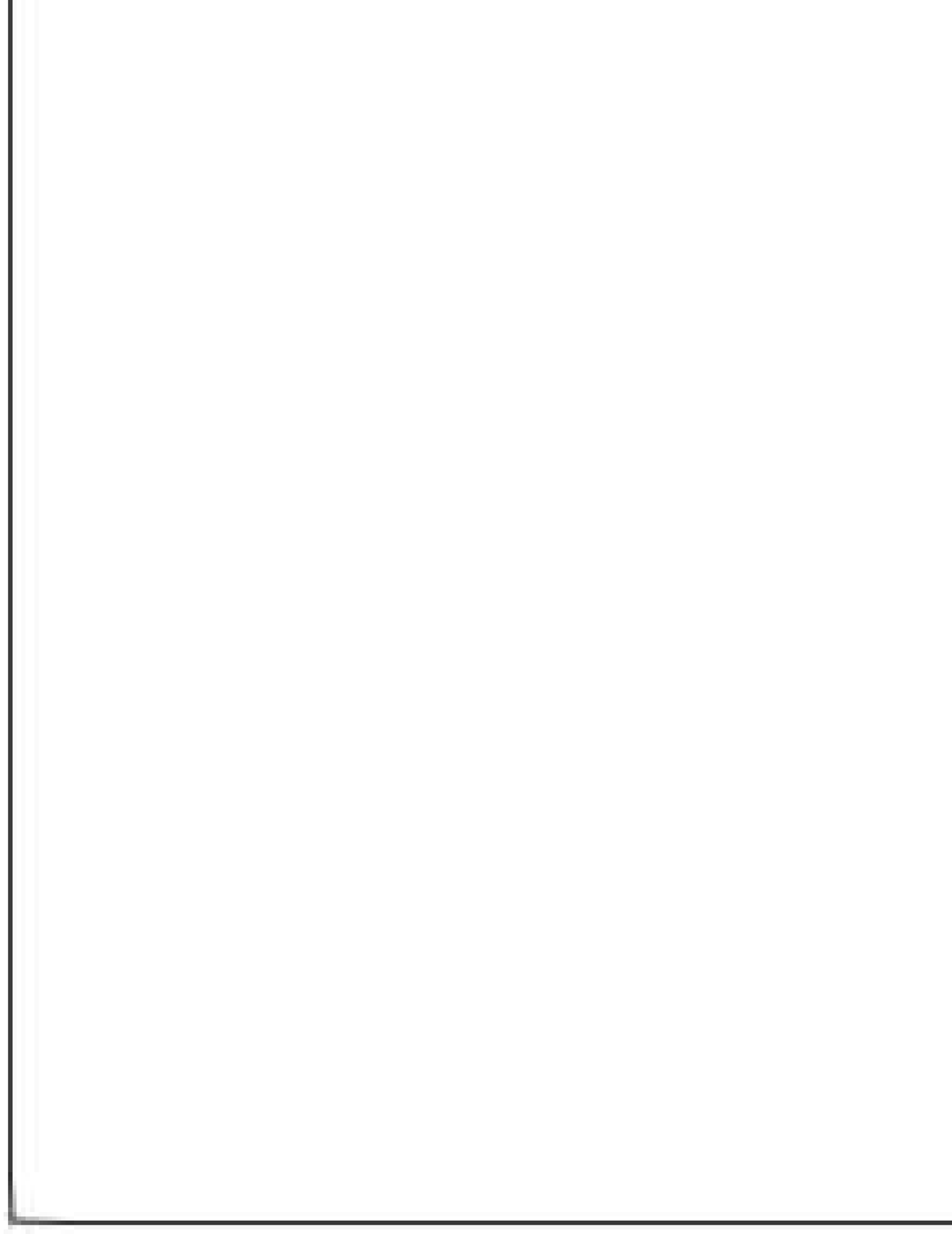} 
	\includegraphics[width=0.2\textwidth]{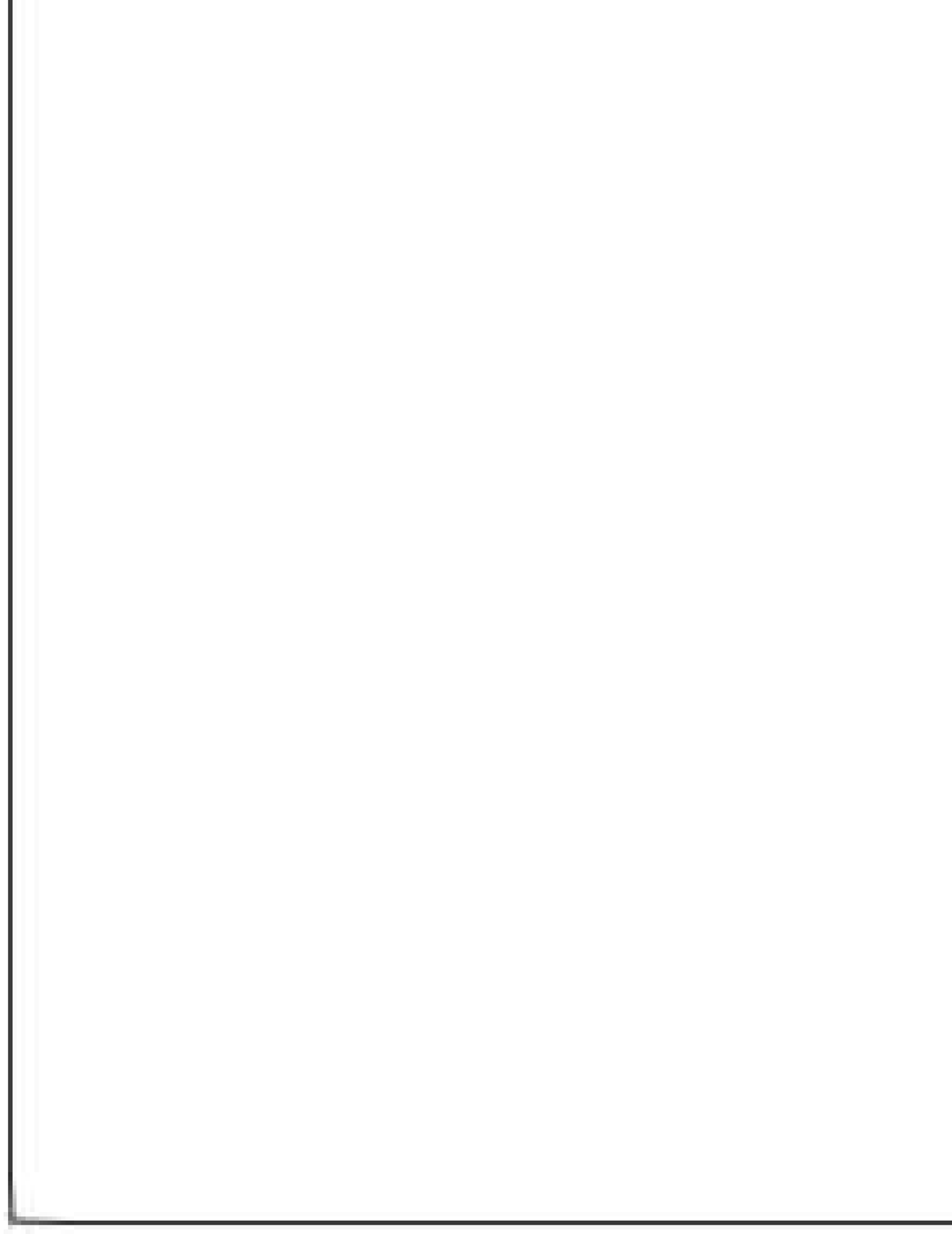} 
	\includegraphics[width=0.2\textwidth]{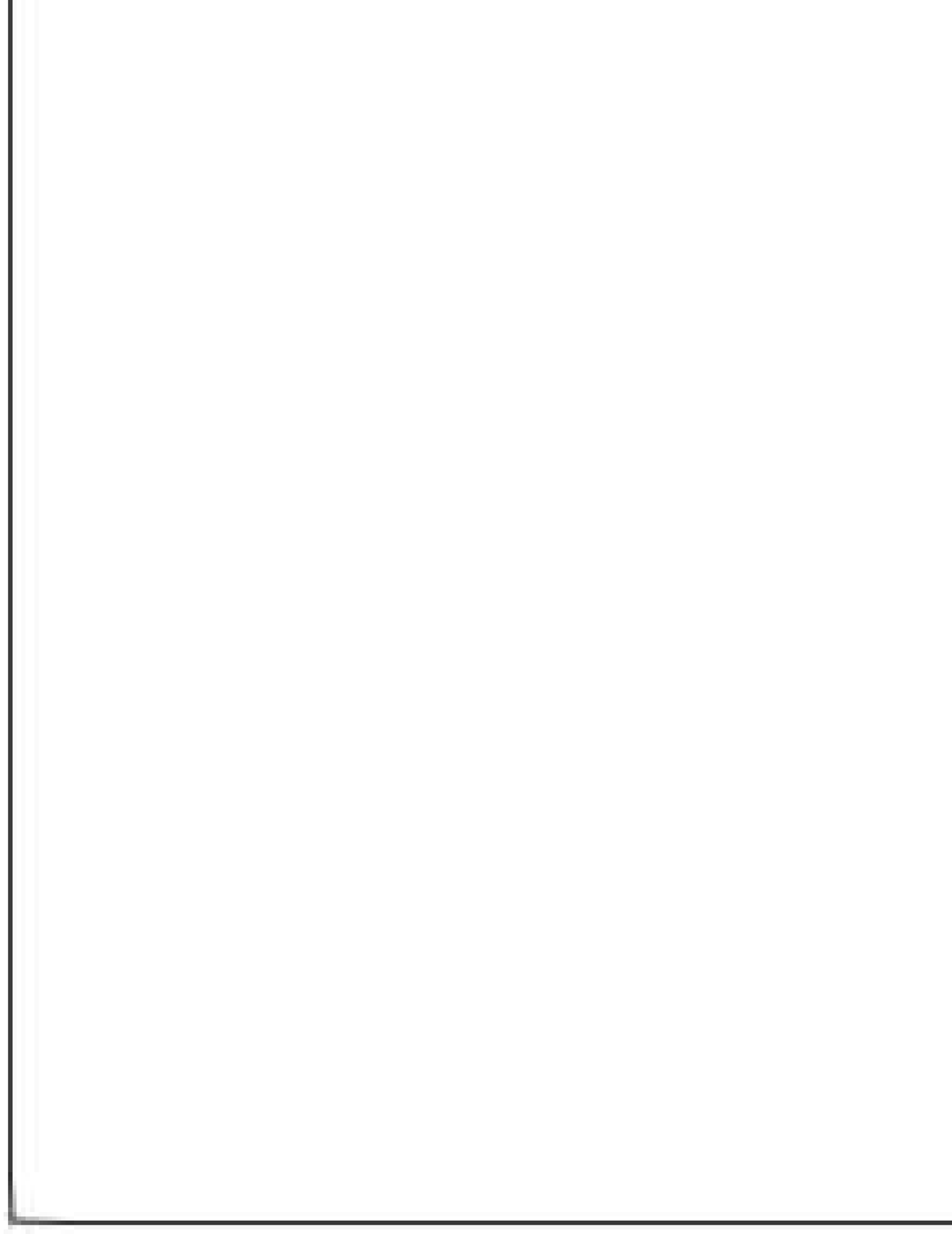} 
	\includegraphics[width=0.2\textwidth]{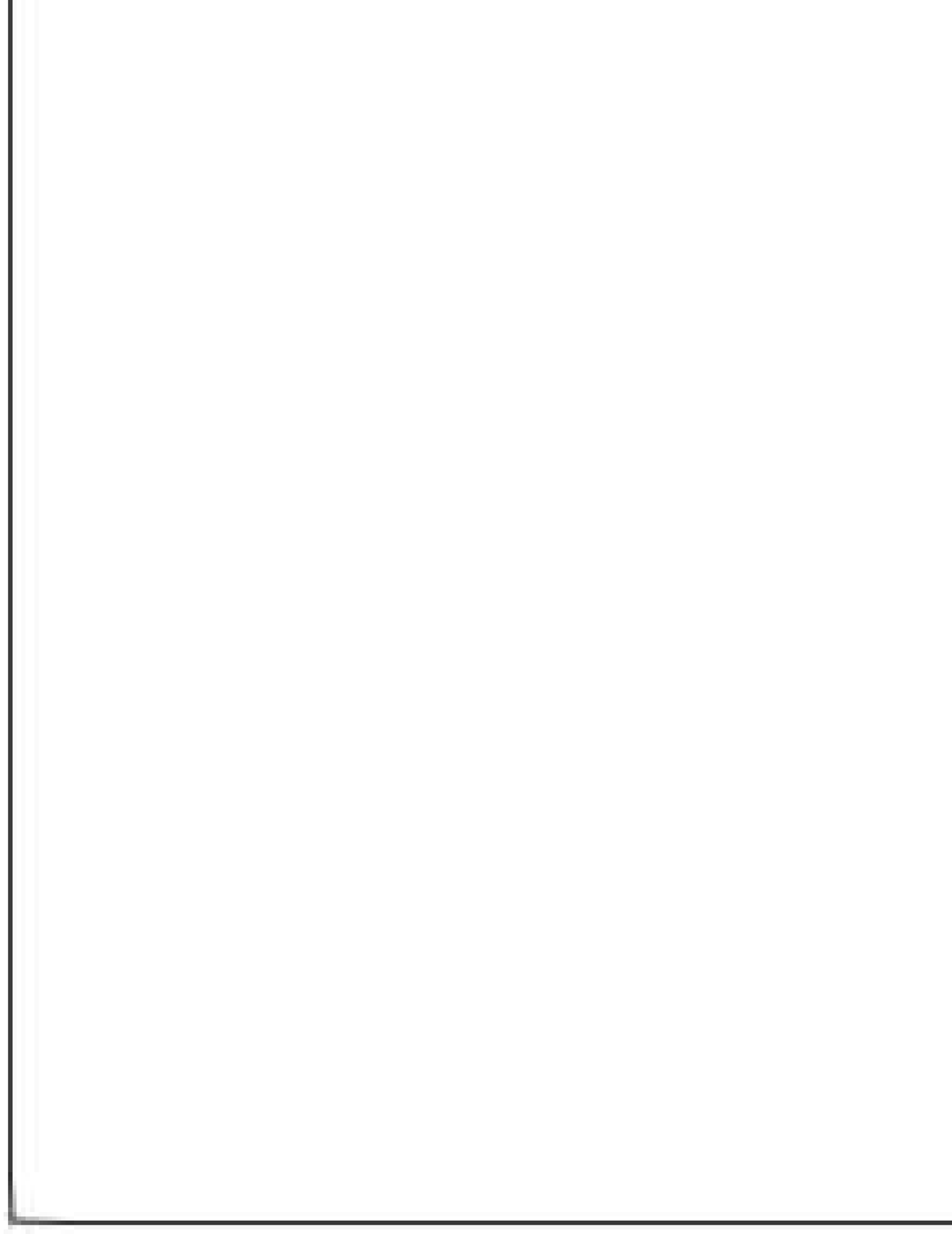}
	\put(-400,47){ (\textit{c})}
	\put(-382,47){\small $\mathit{Fo}=0.0$} 	
	\put(-286,47){\small $\mathit{Fo}=0.63$} 	 	
	\put(-190,47){\small $\mathit{Fo}=1.25$} 	 	
	\put(-93,47){\small $\mathit{Fo}=1.97$} 		 	
	\vspace{0.2cm}
	
	\includegraphics[width=0.2\textwidth]{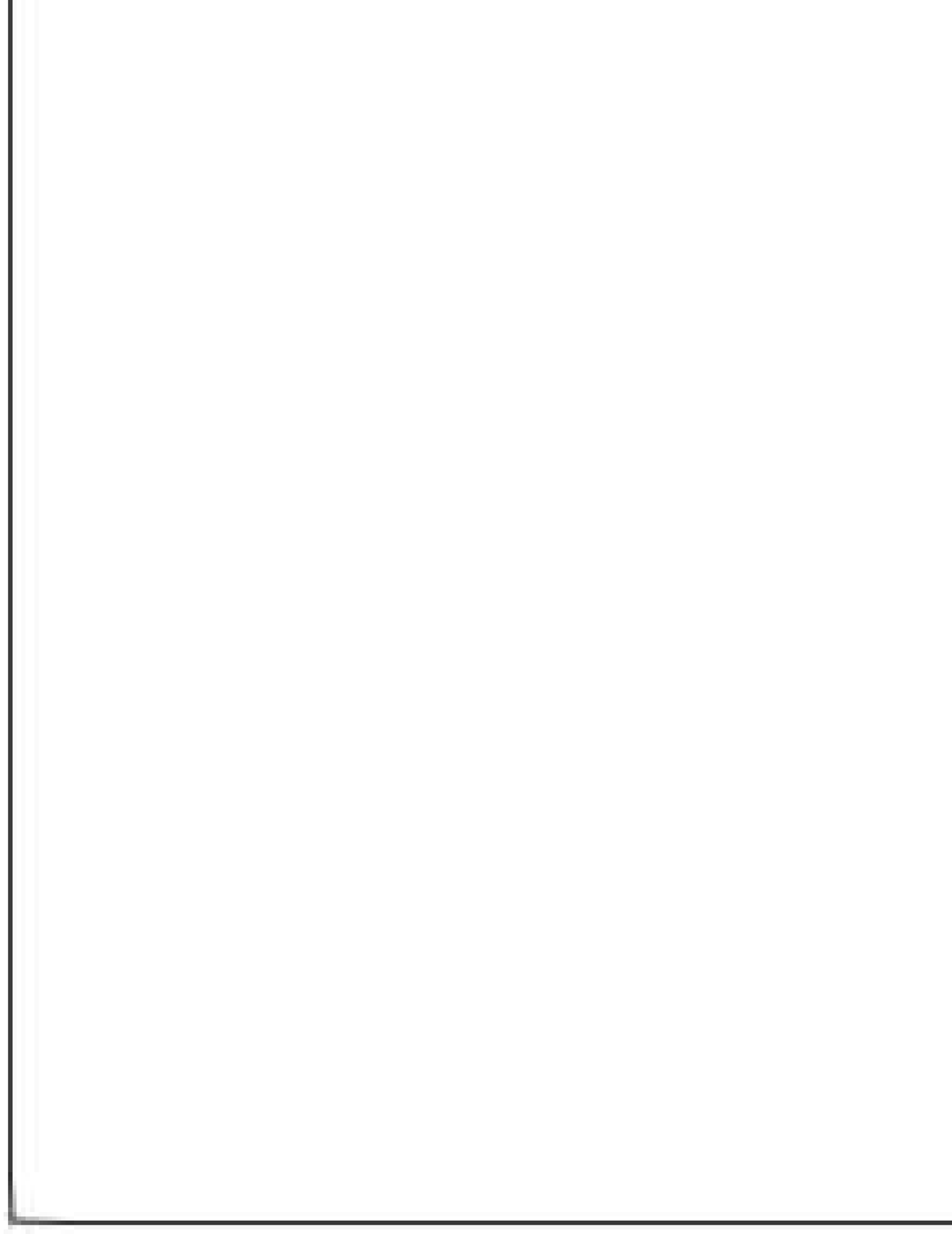} 
	\includegraphics[width=0.2\textwidth]{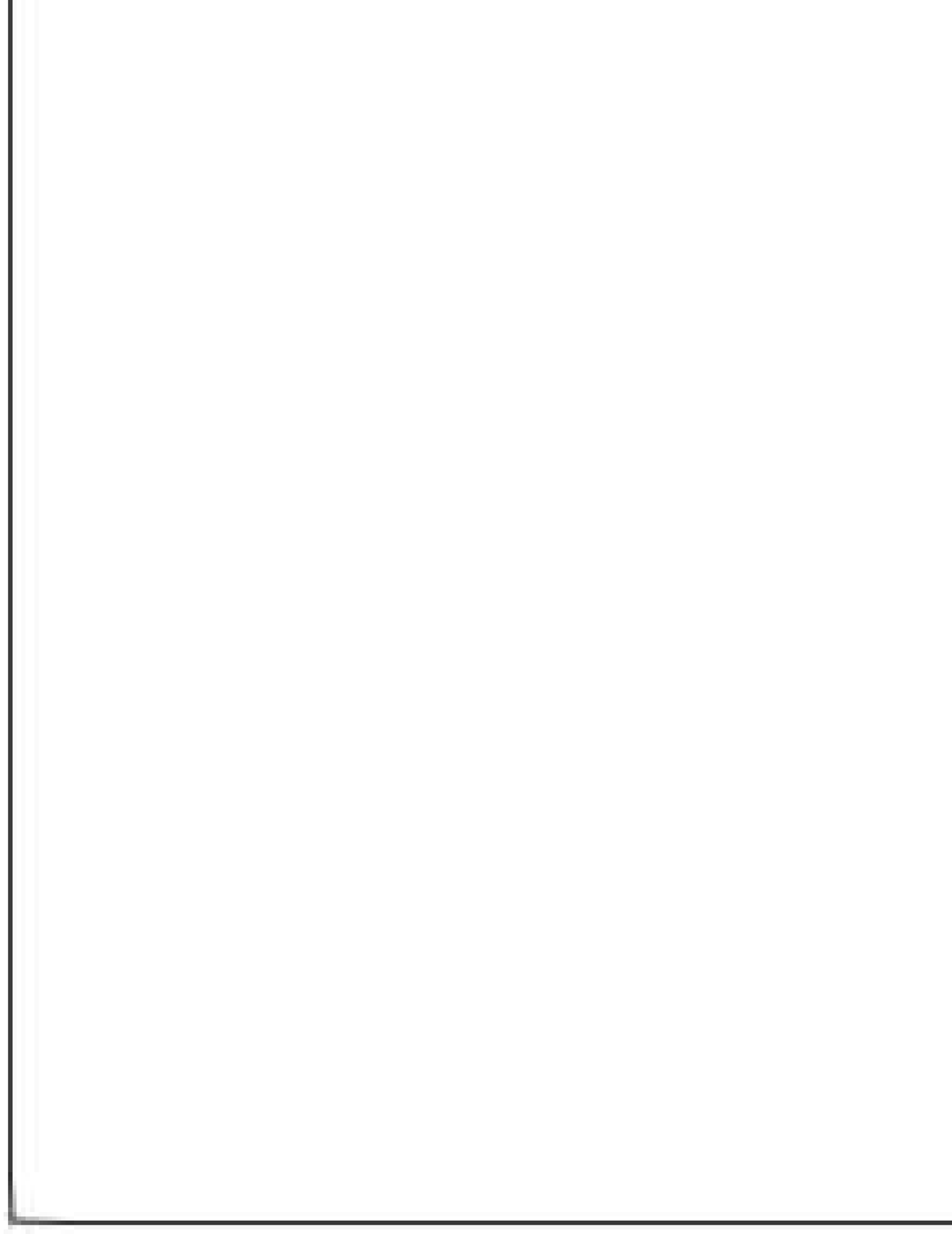} 
	\includegraphics[width=0.2\textwidth]{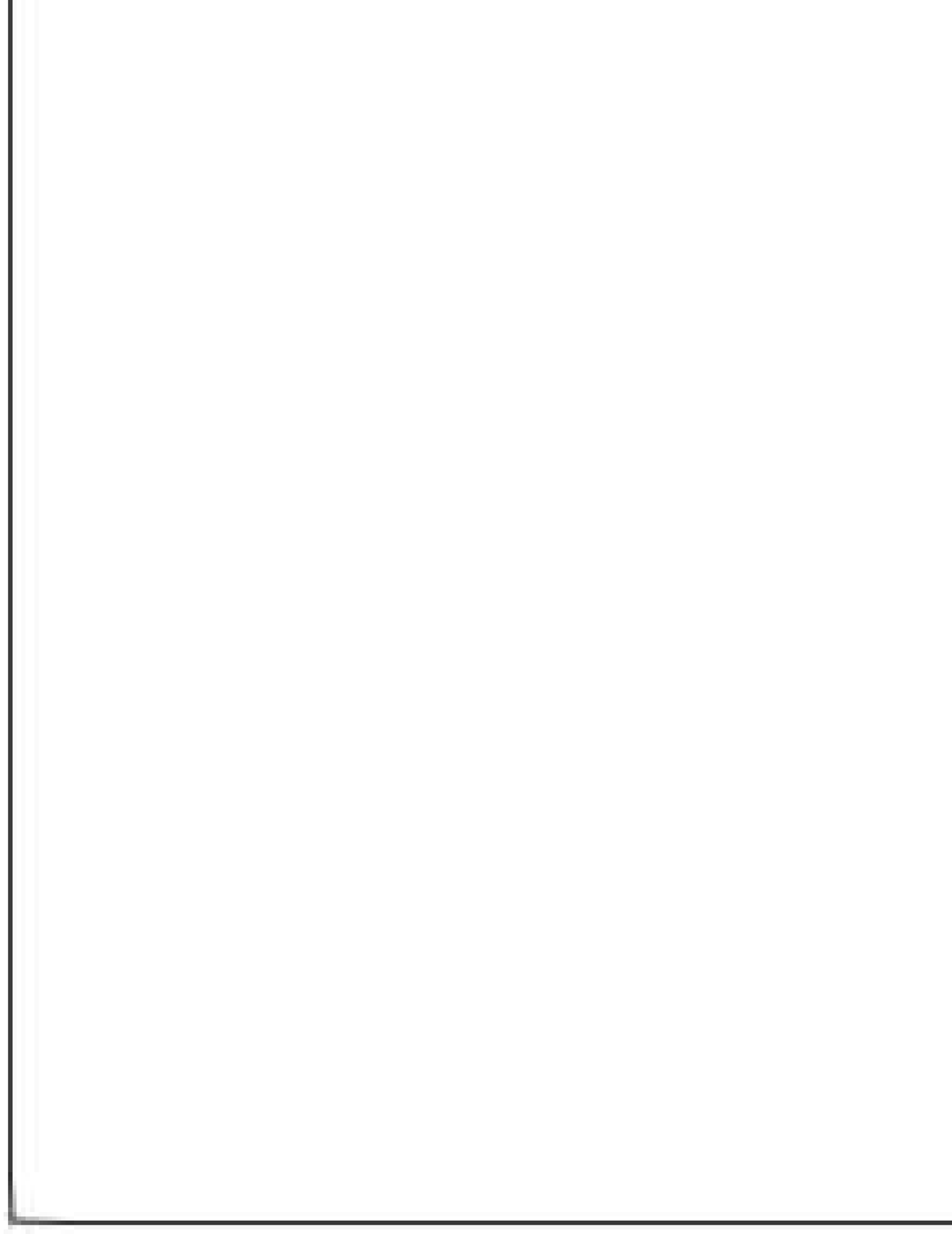} 
	\includegraphics[width=0.2\textwidth]{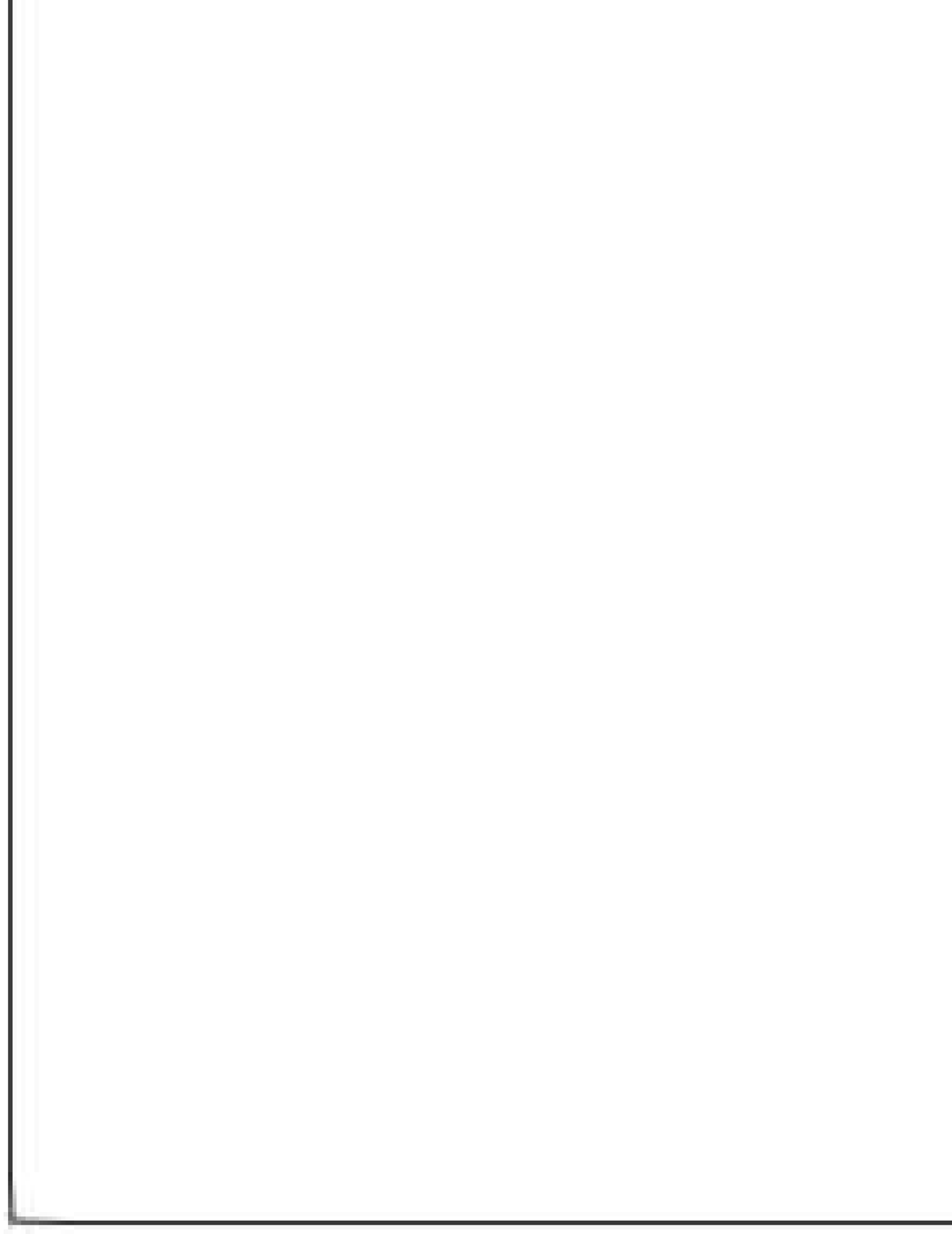}
	\put(-400,47){ (\textit{d})}
	\put(-382,47){\small $\mathit{Fo}=0.0$} 	
	\put(-286,47){\small $\mathit{Fo}=0.63$} 	 	
	\put(-190,47){\small $\mathit{Fo}=1.25$} 	 	
	\put(-93,47){\small $\mathit{Fo}=1.97$} 		 	
	\vspace{0.2cm}

	\caption{ Evolutions of freezing front (solid line) and droplet profiles for $Ca_E=0.0$ (a), $PR^+:R=2.0, S=0.5, Ca_E=1.0$ (b),  $OB^-:R=2.0, S=5.0, Ca_E=1.0$ (c), and $PR^-:R=5.0, S=9.5, Ca_E=1.0$ (d). Other parameters are set as $\mathit{Ste}=0.2, \rho_s/\rho_l=1.1, \alpha_l/\alpha_g=0.2, \theta=120^{\circ}$. }
	\label{fig8}
\end{figure}

We first consider the effect of the electric field on the dynamics of the freezing droplet, and show the evolutions of the freezing front and droplet profile of four cases with different solid-liquid density ratios $\rho_s/ \rho_l $. For all cases, the freezing front (i.e., the isothermal surface at $T=T_m$) gradually moves from the bottom to the top part once the freezing begins. Due to the high thermal diffusivity of the environment, the surrounding fluid cools down quickly, and the freezing front curves upward at the edges of the droplet while remaining flat at the center of the droplet. This phenomenon is also called secondary solidification, as reported by Omid et al. \cite{MohammadipourJFM2024}. Moreover, for the case of $\rho_s/ \rho_l=0.9$, the gas-liquid interface gradually expands outward and eventually forms a conical tip at the top of the droplet due to the influence of surface tension and the solid-liquid density difference ($\mathit{Fo}=2.78$ in Fig. \ref{fig7}, $\mathit{Fo} = \alpha t  /R_0^2$ is the Fourier number). However, for $\rho_s/ \rho_l = 1.1$, the liquid-gas interface gradually contracts inward, and a platform eventually forms at the top of the droplet  ($\mathit{Fo}=1.97$ in Fig. \ref{fig7} ). In addition, the results in Figs. \ref{fig7} and \ref{fig8} also indicate that the shape of the droplet can change significantly once an electric field is applied, compared to the case without the electric field. Actually, under the action of the electric field, the prolate droplet is stretched in the direction of the electric field, and in contrast, the oblate droplet is stretched in a direction perpendicular to the electric field.

\begin{figure}[H]
	\centering
	\includegraphics[width=0.3\textwidth]{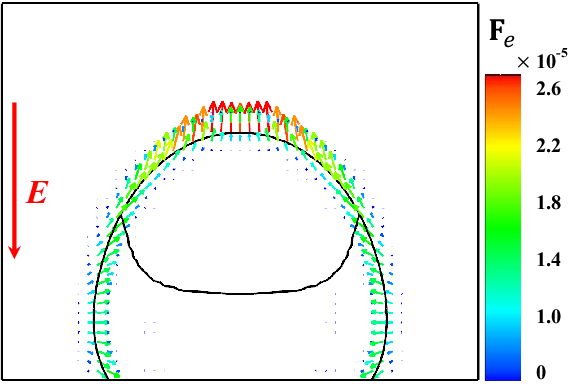} 
	\put(-152,85){\small (\textit{a})}
	\label{fig8a}
	\quad
	\quad
	\includegraphics[width=0.3\textwidth]{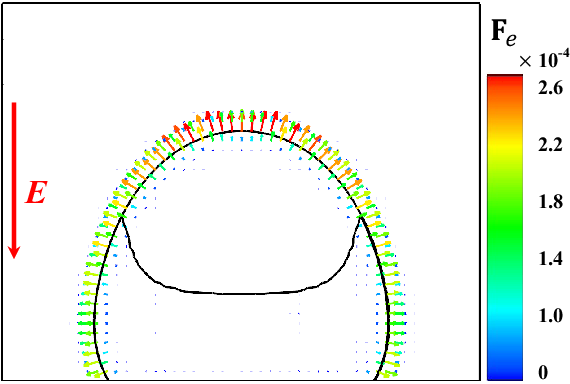}
	\put(-152,85){\small (\textit{b})}	 
	\label{fig8b}
	\quad
	\quad
	\includegraphics[width=0.3\textwidth]{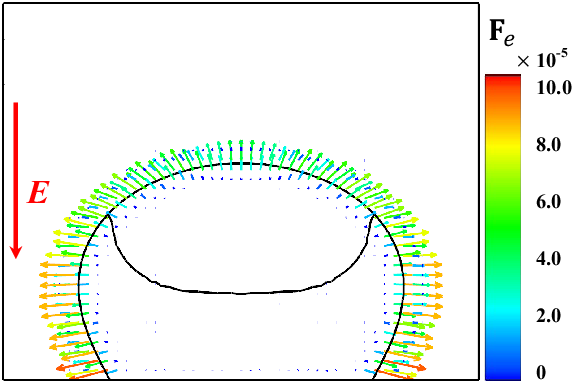} 
	\put(-152,85){\small (\textit{c})}
	\label{fig8c}
	
	\caption{The distribution of electric force at the interface of the freezing droplet $PR^+:R=2.0, S=0.5, Ca_E=1.0$ (a), $OB^-:R=2.0, S=5.0, Ca_E=1.0$ (b), and $PR^-:R=5.0, S=9.5, Ca_E=1.0$ (c) at $\mathit{Fo}=0.63$. Other parameters are set as $\mathit{Ste}=0.2, \rho_s/\rho_l=0.9, \alpha_l/\alpha_g=0.2$ and $\theta=120^{\circ}$.}
	\label{fig9}
\end{figure}

To better understand the effect of the electric force on the droplet shape, Fig. \ref{fig9} shows the electric force acting on the droplet. Due to the accumulation of polarization charges only occurring at the interface, the electric field force only acts on the droplet interface, and it is zero in other regions. For the cases of $PR^+$ and $PR^-$, the droplet poles have a maximum upward electric force, while the droplet equator has inward and outward electric forces. Therefore, the prolate droplet is stretched in the direction of the electric field. For the case of $OB$, the tip of the droplet has a maximum outward electric force, while the droplet poles have a smaller upward electric force. As a result, the oblate droplet is stretched in the horizontal direction under the action of the electric force.

\begin{figure}[H]
	\centering
	\includegraphics[width=0.45\textwidth]{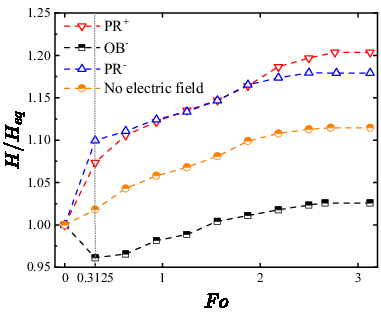} 
	\label{fig10a}
	\put(-210 ,170){\small (\textit{a})}
	\quad
	\includegraphics[width=0.45\textwidth]{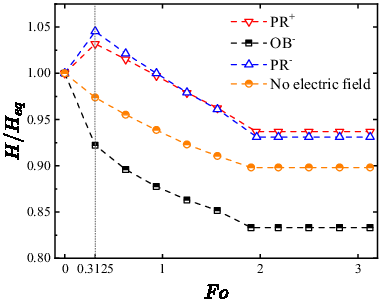} 
	\label{fig10b}
	\put(-210,170){\small (\textit{b})}
	
	\caption{Evolution of droplet height $H/H_{eq}$ with the dimensionless time $\mathit{Fo}$ at $\rho_s/\rho_l=0.9$ (a) and $\rho_s/\rho_l=1.1$ (b). $H_{eq}$ is the initial height of the droplet without an external electric field.}
	\label{fig10}
\end{figure}

We now turn our attention to the effect of the electric field on the final droplet shape after freezing. We present the evolution of the normalized droplet height $H/H_{eq}$ with the dimensionless time $\mathit{Fo}$ at different solid-liquid density ratios in Fig. \ref{fig10}. From this figure, it can be found that for the case without the application of an electric field, the droplet height gradually increases until the droplet completely freezes, which is caused by the volume expansion for the case of $\rho_s/\rho_l=0.9$, while for $\rho_s/\rho_l=1.1$, the droplet height gradually decreases to a steady value due to the volume contraction. Once the electric field is applied, the force induced by the vertical electric field significantly increases the height of the prolate droplet, while the force induced by the horizontal electric field causes the height of the oblate droplet to decrease. Under the effect of the electric field, the normalized height $H/H_{eq}$ of the droplet decreases by $7.53\%$ for $\rho_s/\rho_l=1.1$, while the droplet height increases by $7.68\%$ for $\rho_s/\rho_l=0.9$. Based on the results in Fig. \ref{fig10}, it is also found that in the initial stage ($\mathit{Fo}<0.3125$), the effect of electric field is more significant, while as the droplet is gradually frozen, the effect of electric field seems much weaker. In addition, It is worth noting that for the case of $OB^-$ with $\rho_s/\rho_l=0.9$, the height of the droplet decreases significantly in the initial stage under the horizontal electric field, while as the solid phase increases, the effect of the electric field is weakened, and the droplet expands in volume which leads to a gradual increase in the droplet height.

\subsection{Effect of electric field on heat transfer}

\begin{figure}[H]
	\centering
	\includegraphics[width=0.4\textwidth]{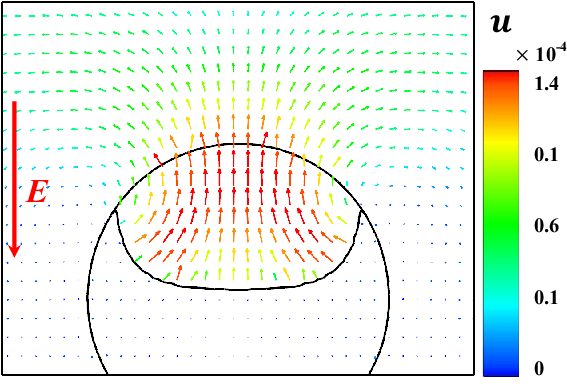} 
	\put(-200,120){\small (\textit{a})}
	\label{fig11a}
	\quad
	\quad
	\includegraphics[width=0.4\textwidth]{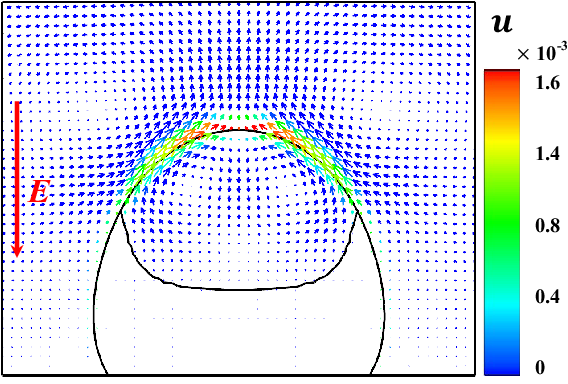}
	\put(-200,120){\small (\textit{b})}	 
	\label{fig11b}

	\includegraphics[width=0.4\textwidth]{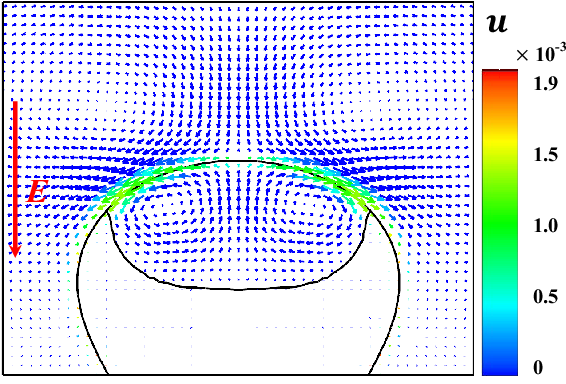} 
	\put(-200,120){\small (\textit{c})}
	\label{fig11c}
	\quad
	\quad
	\includegraphics[width=0.4\textwidth]{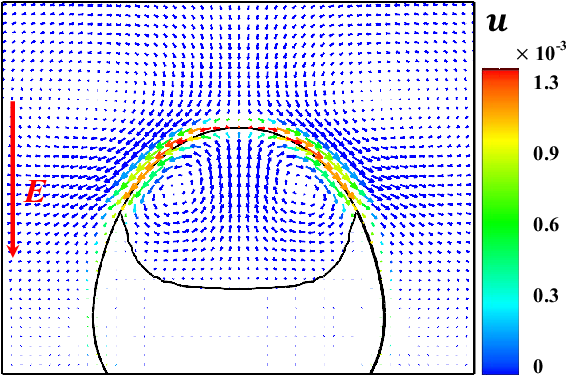} 
	\put(-200,120){\small (\textit{c})}
	\label{fig11d}

	\caption{The distributions of electric force at the interface of the freezing droplet and flow fields of different cases. $Ca_E=0.0$ (a), $PR^+:R=2.0, S=0.5, Ca_E=1.0$ (b), $OB^-:R=2.0, S=5.0, Ca_E=1.0$ (c), and $PR^-:R=5.0, S=9.5, Ca_E=1.0$ (d) at $Fo=0.63$. Other parameters are $\mathit{Ste}=0.2, \rho_s/\rho_l=0.9, \alpha_l/\alpha_g=0.2, \theta=120^{\circ}$.}
	\label{fig11}
\end{figure}

In this part, we will focus on the influence of the electric field on heat transfer during the droplet freezing process. Under the action of the electric field, the droplet deforms, and also some symmetric vortexes both inside and outside the droplet are formed. Fig. \ref{fig11} shows the velocity field inside and outside the droplet at $\mathit{Fo} = 0.63$. In the absence of an electric field, the freezing front continuously rises, and the internal and external velocities of the droplet are upward; in this case, the flow is weak, and no vortex is formed. When the electric field is applied, the flow strength is enhanced, and two pairs of symmetric vortexes are formed inside and outside the droplet. Specifically, for the $PR^+$ mode, the internal vortexes flow from the poles to the equator, and the external vortexes flow from the equator to the poles. In contrast, for the $OB^-$ and $PR^-$ modes, the internal vortexes flow from the equator to the poles, and the external vortexes flow from the poles to the equator. When an electric field is applied, besides the change of flow field, the convective heat transfer is also enhanced.

To quantify the contributions of convection and conduction to heat transfer, here we use the Peclet number $(\mathit{P e=D_d \bar{u}_d / \alpha})$ to characterize the convective strength induced by the electric field inside the droplet, where $D_d$, $\bar{u}_d$, and $\alpha$ represent the characteristic length, the average flow rate, and thermal diffusivity of the droplet, respectively.

\begin{figure}[H]
	\centering
	\includegraphics[width=0.45\textwidth]{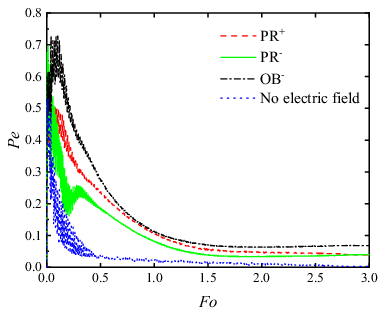} 
	\label{fig12a}
	\put(-210 ,170){\small (\textit{a})}
	\quad
	\includegraphics[width=0.45\textwidth]{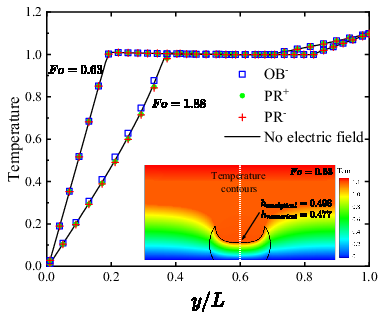} 
	\label{fig12b}
	\put(-210,170){\small (\textit{b})}
	
	\caption{ The evolution of Pe with the dimensionless time $\mathit{Fo}$ (a), the temperature distributions along the vertical centerline of a domain at different moments (b).}
	\label{fig12}
\end{figure}

We present the variation of $\mathit{Pe}$ with the dimensionless time $\mathit{Fo}$ for different cases in Fig. \ref{fig12}(a) and find that $Pe$ increases significantly after applying the electric field due to the enhancement of the convection. However, the fact that $\mathit{Pe}<1$ during the freezing process indicates that convective heat transfer is much weaker than thermal conduction, and thermal conduction becomes the primary mode of heat transfer. This can be further confirmed by the temperature distribution along the droplet centerline, as shown in Fig. \ref{fig12}(b). It can also be observed that the temperature distribution of the droplet does not change after applying the electric field, which is due to the fact that the electric field does not affect thermal conduction and has a limited impact on convective heat transfer at $\mathit{Ste}=0.2$. Additionally, the position of the freezing front in the one-dimensional Stefan problem follows the relation $h=\sqrt{\mathit{2 Ste Fo}}$ \cite{WatanabeCES1992}. We measure the numerical and analytical values ($h_{numerical}=0.477$ and $h_{analytical}=0.498$), and the relative error is about 4.3\%, indicating that the one-dimensional solution can still provide a good approximation for the droplet's solidification at the initial stages.

\begin{figure}[H]
	\centering
	\includegraphics[width=0.45\textwidth]{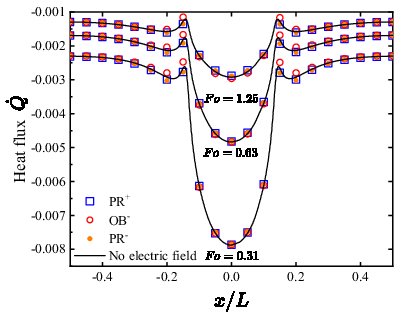} 
	\label{fig13a}
	\put(-210 ,170){\small (\textit{a})}
	\quad
	\includegraphics[width=0.45\textwidth]{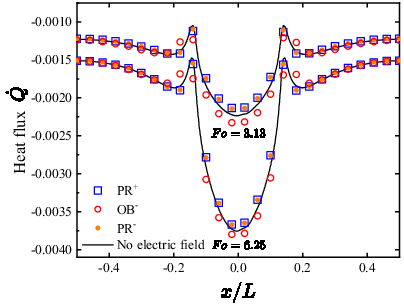} 
	\label{fig13b}
	\put(-210,170){\small (\textit{b})}
	
	\caption{ The distribution of heat flux $\dot{Q}$ along the cold substrate for different values of $\mathit{Ste}$ (a) $\mathit{Ste}=0.2$ and (b) $\mathit{Ste}=0.05$.}
	\label{fig13}
\end{figure}

The above discussion indicates that heat conduction is dominant in the electro-freezing process wetting droplet. Fig. \ref{fig13} illustrates the evolution of local heat flux on the substrate with the dimensionless time $\mathit{Fo}$, here the local heat flux is calculated by using Fourier's law $\dot{Q}=-\lambda(\partial T / \partial y)_{y=0}$, in which the local temperature gradient at the substrate can be calculated using a second-order finite difference scheme $(\partial T / \partial y)_{y=0}=\left(4 T_{y=1}-T_{y=2}-3 T_{y=0}\right) /(2 \Delta y)$. From Fig. \ref{fig13},  at the initial stage ($\mathit{Fo}=0.31$), the heat flux is relatively high due to the lower temperature of the substrate than the droplet, resulting in a large temperature gradient and a high rate of heat conduction. As the droplet gradually freezes, the unfrozen portion decreases, reducing the total amount of latent heat releases, and the heat flux correspondingly diminishes. In addition, for the cases with a smaller $\mathit{Ste}$, the heat during the freezing process is primarily used to overcome latent heat rather than sensible heat transfer. Since the release of latent heat requires a substantial amount of energy and the efficiency of sensible heat transfer is lower, the freezing rate of the droplet decreases. Therefore, the droplet freezing time is enlarged for this case, making the effect of the electric field on the droplet more significant. For the $PR^+$ and $PR^-$ modes, the electric field stretches the droplet, shortening the contact line and reducing the heat flux. However, for the $OB^-$ mode, the electric field compresses the droplet, extending the contact line and increasing the heat flux. As the $\mathit{Ste}$ increases, the latent heat decreases, reducing the energy required for droplet freezing and thus increasing the freezing rate. The effect of the electric field on the freezing process is weakened by the rapid freezing of the droplet into the solid. 


\subsection{Effect of electric field on freezing time}

\begin{figure}[H]
	\centering
	\includegraphics[width=0.45\textwidth]{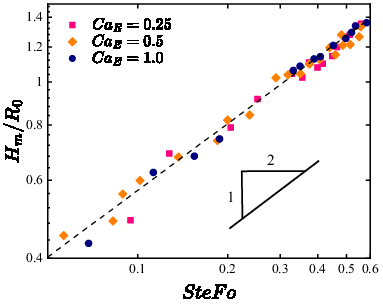} 
	\label{fig14a}
	\put(-210 ,170){\small (\textit{a})}
	\quad
	\includegraphics[width=0.45\textwidth]{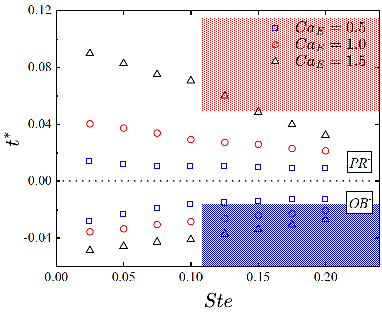} 
	\label{fig14b}
	\put(-210,170){\small (\textit{b})}
	
	\caption{ (a) The scaling relationship between the final dimensionless height of droplets $H_m/R_0$ and $SteFo$. (b) The freezing time at different electric field strengths, the dimensionless freezing time $t^*$ is scaled by $t^*=(t_{Ca_E}-t_{Ca_E=0})/t_{Ca_E=0}$, where $t_{Ca_E}$ and $t_{Ca_E=0}$ are the freezing time with and without electric field, respectively.}
	\label{fig14}
\end{figure}

Finally, we consider the effect of the electric field on the freezing time, which represents the time required for the droplet to completely freeze. Based on the one-dimensional solution of Stefan solidification solution, the freezing time $t_{\mathit{fre}}$ is related to the final height $H_m$ of the droplet $H_m=2 \hat{\lambda} \sqrt{\alpha t_{\mathit{fre}}}$, where $\hat{\lambda}$ is the root of the equation for Stefan problem. Note that when the contribution of latent heat $L$ is much larger than that of the sensible heat contribution, i.e.,$\mathit{Ste}<1$, it yields $\hat{\lambda}=\sqrt{c_i \Delta T / 2 L}$. By introducing the Stefan number $Ste$ and the Fourier number $Fo$ to the above equation, the following relationship between droplet final height and freezing time is obtained.
\begin{equation}
	H_m / R_0 \propto \mathit{Ste}^{1 / 2} \mathit{Fo}^{1 / 2}.
\end{equation}
Fig. \ref{fig14}(a) plots the evolution of the final droplet height with $\mathit{SteFo}$, and the current numerical results are consistent with the above power-law expression. In addition, it can also be seen that the droplet freezing time increases with the final height of the droplet.

To illustrate the effect of the electric field on droplet freezing time, Fig. \ref{fig14}(b) shows the variation of droplet freezing time with $\mathit{Ste}$ under different electric field strengths, where the dimensionless evaporation time is scaled by $t^*=(t_{Ca_E}-t_{Ca_E=0})/t_{Ca_E=0}$. Since the trends for the $PR^+$ and $PR^-$ cases are similar, only the results of the $PR^-$ case are shown in this figure. We can find that for the $OB^-$ case, the electric field enhances the freezing, which leads to a shorter freezing time. On the contrary, for the case of $PR^-$, the electric field suppresses the freezing and thus prolongs the freezing time. As discussed previously, when an electric field is applied, the electric force compresses the oblate droplet, causing its height to decrease and the contact line to increase, which enhances heat transfer and reduces thermal resistance. In contrast, the prolate droplet is stretched under the effect of the electric field, causing its height to increase and the contact line to decrease, which also enhances heat transfer and reduces thermal resistance. Therefore, the electric field shortens the freezing time for oblate droplets while prolonging it for prolate droplets. Additionally, the results also indicate that as $\mathit{Ste}$ increases, the effect of the electric field on droplet freezing time becomes weak. Due to the gradual decrease in droplet freezing time under high Stefan number conditions, the duration of the electric field's influence is significantly reduced.

\section{Conclusions}
\label{sec6}

In this work, we first developed a phase-field LB method to simulate the electro-freezing process in gas-liquid-solid systems, considering the volume expansion or contraction due to the density difference between liquid and solid. The numerical method is first tested by the three-phase Stefan problem, freezing on a cold wall surface, and droplet deformation under an electric field. Subsequently, we investigated the freezing dynamics of wetting droplets in a uniform electric field, focusing on the effects of the electric field on droplet morphology evolution, heat transfer, and freezing time.

The results indicate that the prolate droplet is stretched along the direction of the field, resulting in an increased droplet height and a shortened contact line. Conversely, the oblate droplet is compressed by the electric field, leading to a decreased droplet height and an elongated contact line. Although the electric field enhances the flow inside the droplet, the convection is limited by freezing, so thermal conduction remains the primary mode of heat transfer. The time evolution of the freezing front under the electric field follows the classic power law $(t^{0.5})$, indicating that the electric field mainly affects droplet freezing by changing the droplet shape. Compared to the case without an electric field, the freezing time of prolate droplets is longer due to the high thermal resistance and low heat transfer area of the electric lift mode. In contrast, the freezing time of oblate droplets is shortened due to the low thermal resistance and high heat transfer area of the electric squeeze mode.

\section*{Acknowledgments}
This research was supported by the National Natural Science Foundation of China (Grants No. 12072127 and No. 123B2018), the Interdisciplinary Research Program of HUST (2024JCYJ001 and 2023JCYJ002), and the Fundamental Research Funds for the Central Universities, HUST (No. YCJJ20241101 and No. 2024JYCXJJ016). The computation was completed on the HPC Platform of Huazhong University of Science and Technology.


\end{document}